\documentclass[10pt,journal,compsoc]{IEEEtran}
\ifCLASSOPTIONcompsoc
  \usepackage[nocompress]{cite}
\else
  \usepackage{cite}
\fi

\usepackage{times}
\usepackage[mathletters]{ucs}
\usepackage[utf8x]{inputenc}
\usepackage{epsfig}
\usepackage{amsmath}
\usepackage{bm}
\usepackage{amssymb}
\usepackage{mathabx}

\usepackage{mathrsfs}
\usepackage{caption}
\usepackage{pifont}
\usepackage{wasysym}
\usepackage{array}
\usepackage{caption}
\usepackage{color}
\usepackage{xcolor}
\usepackage{colortbl}
\usepackage{hyperref}
\usepackage{wrapfig}
\usepackage{multicol}
\usepackage{hhline}
\usepackage{booktabs} 
\usepackage{multirow}
\usepackage{color}
\usepackage{xcolor}
\usepackage{colortbl}
\usepackage{wrapfig}
\newcommand{\xmark}{\ding{55}}%
\newcommand{\eg}{\textit{e}.\textit{g}.}
\newcommand{\etal}{\textit{et al}.}
\newcommand{\ie}{\textit{i}.\textit{e}.}
\usepackage{multirow}
\usepackage{makecell, multirow, tabularx}
\usepackage{adjustbox}
\usepackage{tabularx}
\usepackage{xspace}
\usepackage{booktabs}

\usepackage{pdfpages} 
\usepackage{pgffor} 
\usepackage{xr} 

\makeatletter
\AtBeginDocument{\let\LS@rot\@undefined}
\makeatother

\def\supplementfilename{supplement}

\pdfximage{\supplementfilename.pdf}
\def\numbersupplementpages{\the\pdflastximagepages}

\newif\ifarXiv
\arXivtrue 

\ifCLASSINFOpdf
\else
\fi
\begin{document}
%
\title{Deep Learning for HDR Imaging: State-of-the-Art and Future Trends}
%
%
%
%

\author{Lin Wang,~\IEEEmembership{Student Member,~IEEE,}
        and~Kuk-Jin Yoon,~\IEEEmembership{Member,~IEEE}
   
\IEEEcompsocitemizethanks{\IEEEcompsocthanksitem L. Wang and K.-J. Yoon are with the Visual Intelligence Lab, Department of Mechanical Engineering, Korea Advanced Institute of Science and Technology, 291 Daehak-ro, Guseong-dong, Yuseong-gu, Daejeon 34141, Republic of Korea. 
E-mail: \{wanglin, kjyoon\}@kaist.ac.kr.}
\thanks{Manuscript received May 15, 2021; revised August 26, 2021.\hfil\break(Corresponding author: Kuk-Jin Yoon)}}

%
%

\markboth{Journal of \LaTeX\ Class Files,~Vol.~14, No.~8, August~2015}%
{Shell \MakeLowercase{\textit{et al.}}: Bare Demo of IEEEtran.cls for Computer Society Journals}
%



\IEEEtitleabstractindextext{%
\begin{abstract}
High dynamic range (HDR) imaging is a technique that allows an extensive dynamic range of exposures, which is important in image processing, computer graphics, and computer vision. In recent years, there has been a significant advancement in HDR imaging using deep learning (DL). This study conducts a comprehensive and insightful survey and analysis of recent developments in deep HDR imaging methodologies. 
We hierarchically and structurally group existing deep 
HDR imaging methods into five categories based on (1) number/domain of input exposures, (2) number of learning tasks, (3) novel sensor data, (4) novel learning strategies, and (5) applications.  Importantly, we provide a constructive discussion on each category regarding its potential and challenges. Moreover, we review some crucial aspects of deep HDR imaging, such as datasets and evaluation metrics. Finally, we highlight some open problems and point out future research directions.
\end{abstract}

}

\maketitle

\IEEEdisplaynontitleabstractindextext

%
\IEEEpeerreviewmaketitle

\vspace{-10pt}
\IEEEraisesectionheading{\section{Introduction}\label{sec:introduction}}
%
%
%
%
\vspace{-5pt}
\IEEEPARstart{H}{igh} dynamic range (HDR) imaging, an important field in image processing, computer graphics/vision, and photography, is a technique that allows a greater dynamic range of exposures than traditional imaging techniques. 
It aims to accurately represent a wide range of intensity levels captured in real scenes, ranging from sunlight to shadows \cite{HDRI_wiki,HDRI_github}. 
Using HDR imaging, real-world lighting can be captured, stored, transmitted, and fully used in various applications without the need to linearize the signal and handle clamped values \cite{banterle2017advanced}.
Conventional HDR imaging mainly uses special HDR cameras to capture HDR images \cite{tiwari2015review,tursun2015state,johnson2015high}; however, these cameras are prohibitively expensive for general users. An alternative is to create HDR content from virtual environments using 
rendering tools. However, this approach
is mostly explored in the entertainment industry, \eg, gaming and virtual reality (VR) \cite{wang2019co,tiwari2015review,banterle2017advanced}. 

In addition to the aforementioned approaches, a common method is to reconstruct HDR images from the visual content captured by low-dynamic-range (LDR) cameras using specially designed algorithms. Among these algorithms, there are two commonly explored methods. 
The first is to generate HDR content by fusing multiple LDR images of the same scene at different exposure times~\cite{yan2020deep,banterle2017advanced,kalantari2017deep}. However, as capturing LDR contents at different exposures requires the use of certain software/hardware technology, it is often difficult to create datasets. Therefore, the second approach is to generate HDR content from a single-exposure image~\cite{eilertsen2017hdr,lee2018recursive,khan2019fhdr,ning2018learning,liu2020single}. 

Deep learning (DL) has recently been applied to HDR imaging \cite{zeng2020learning}. DL-based HDR imaging methods often achieve state-of-the-art (SoTA) performances on various benchmark datasets. Deep neural network (DNN) models have been developed based on diverse architectures, ranging from convolutional neural networks (CNNs) \cite{eilertsen2017hdr,zhang2017learning,kalantari2017deep} to generative adversarial networks (GANs)~\cite{ma2019fusiongan,kim2020jsi,yang2020ganfuse}. In general, SoTA-DNN-based methods differ in terms of five major aspects: network design that considers the number and domain of input LDR images \cite{eilertsen2017hdr,liu2020single,kalantari2017deep}, purpose of HDR imaging in multitask learning \cite{kim2020end,kim2018multi}, different sensors being used to obtain deep HDR imaging ~\cite{han2020neuromorphic,wang2019event,kuang2020thermal}, novel learning strategies~\cite{ma2019fusiongan,kumar2017no,pan2021metahdr}, and practical applications \cite{wang2020traffic,mukherjee2020backward,weiher2019domain}.

\begin{figure*}[t!]
    \centering
    \captionsetup{font=small}
    \includegraphics[width=\textwidth]{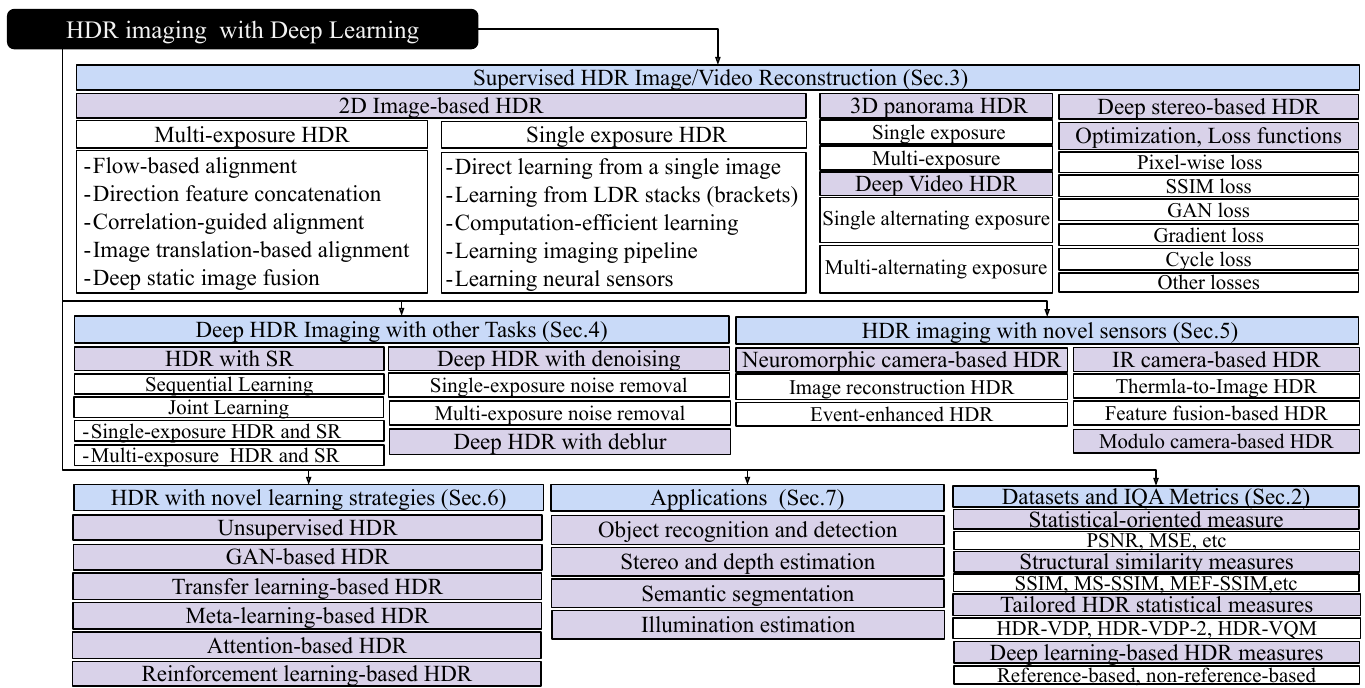}
    \vspace{-12pt}
    \caption{Hierarchical and structural taxonomy of HDR imaging with deep learning.}
    \label{fig:taxomony_hdr}
    \vspace{-5pt}
\end{figure*}

This study provides a comprehensive and systematic overview of recent developments in deep HDR imaging methods. Previous surveys \cite{gallo2016stack,tiwari2015review,srikantha2012ghost,tursun2015state,johnson2015high} focused on conventional HDR imaging algorithms, especially deghosting methods in dynamic scenes \cite{tursun2015state,srikantha2012ghost}, multiple-exposure fusion in the image and radiance domain \cite{tiwari2015review,johnson2015high}, and stack-based algorithms for HDR reconstruction \cite{gallo2016stack}.  Unlike these existing surveys, this study highlights the importance of DL and investigates recent advances in deep HDR imaging, both methodically and comprehensively.   

The main contributions of this study are three-fold: 
(I) We provide a comprehensive overview of deep HDR imaging methods, including the problem definition, datasets, evaluation metrics, a taxonomy, and applications.
(II) We conduct an analytical study of recent trends in deep HDR imaging hierarchically and structurally and offer insights into the opportunities and challenges of each category. 
(III) We discuss the open challenges and problems that HDR imaging faces on the whole and identify future directions to provide guidance to further research in this area.

In the following sections, we discuss and analyze various aspects of recent advances in deep HDR imaging. Fig.~\ref{fig:taxomony_hdr} shows the structural and hierarchical taxonomy used in this study. The remainder of this paper is organized as follows. In Sec.~\ref{sec2}, we define the HDR imaging problem, datasets, and evaluation metrics.  In Sec.~\ref{hdr_recon}, we review and analyze SoTA-supervised methods, including multiple-exposure and single-exposure HDR imaging/video methods. Sec.~\ref{mtl_hdr} focuses on joint HDR imaging with other learning tasks, \eg, image super-resolution. In Sec.~\ref{hdr_novelsensors}, we investigate recent deep HDR imaging methods using novel camera sensors. Sec.~\ref{learning_stratey} covers methods that use novel learning paradigms, \eg, meta-learning. Sec.~\ref{applications} then scrutinizes the applications, followed by Sec.~\ref{sec:discussions}, where we discuss open problems and future directions.     

 

\section{Problem, Datasets, and Evaluation}
\label{sec2}
\subsection{ Definition of Deep HDR Imaging }
The LDR camera usually sets an appropriate exposure time $\Delta t$ and relies on the camera response function (CRF) $f_{CRF}$ to map the irradiance $E$ of scenes to the LDR image $x$ ~\cite{chen2020learning}. 
\begin{equation}
\label{eq1}
   x = f_{CRF} (E\Delta t)
\end{equation}

Fundamentally, deep HDR imaging can be regarded as the process of reconstructing HDR image content from the LDR counterpart(s) using DNNs.
The aim is to learn a mapping function $M$ with parameters $\theta$ that maps a set of multi-exposure LDR images $X=\{x_1,x_2, \cdots, x_n\}$ to an HDR image $y$ \cite{yan2019attention}, where $n$ is the number of LDR images. For single-exposure HDR imaging, the set size is set to one.  Mathematically, this is formulated as follows:
\begin{equation}
\label{eq2}
    Y = M(X; \theta).
\end{equation}

\begin{table*}[t!]
\captionsetup{font=small}
\caption{Summary of publicly available benchmark datasets. N/A: not available.}
\vspace{-15pt}
\small
\begin{center}
\begin{tabular}{c|c|c|c|c|c}
\hline
 Dataset & Amount &  Data type & GT & Resolution & Scene details \\
\hline
\hline
Kalantari \etal \cite{kalantari2017deep} & 89 (train + test) & Real & Yes & 1500 × 1000 & Multi-exposure with 100 dynamic scenes\\
Sen \etal \cite{sen2012robust} & 8 (test only) & Real & No & 1350 × 900 & Multi-exposure with dynamic scenes \\ 
Tursun \etal~\cite{tursun2016objective} & 16 (test only) &  Real & No & 1024 × 682 & Multi-exposure with indoor and outdoor scenes \\ 
Endo \etal~\cite{endoSA2017} & 1043 (train only) &  Synthetic & Yes & 512 × 512 & Single exposure to indoor and outdoor scenes \\
Eilertsen \etal~\cite{eilertsen2017hdr} & 96 (test only) & Synthetic &  Yes &  1024 × 768 & Single exposure to static scenes \\
Zhang \etal~\cite{zhang2017learning} & 41222 (train + test) & Synthetic + Real & Yes & 128 × 64 &  Single-exposure panorama scenes\\
HDREye~\cite{nemoto2015visual} & 46 (test only) & Synthetic & Yes & N/A & Static scenes (both indoor and outdoor) \\
Lee \etal~\cite{lee2018deep} & 96 (train + test) & Synthetic & Yes & 512 × 512 & Static scenes with different exposure levels \\
Prabhakar \etal~\cite{prabhakar2019fast} & 582 (train + test) & Real & Yes &1-4 Megapixels & Dynamic scenes with real-life object motions \\
Cai \etal~\cite{cai2018learning} & 4413 (train + test) & Synthetic & Yes & 3 K-6K & Multi-exposure static indoor and outdoor scenes \\
Liu \etal~\cite{liu2020single} & N/A (train + test) &Synthetic + Real & Yes &  1536 × 1024 & Single exposure to static scenes \\ 
Kim \etal~\cite{kim2019deep} & 59818 frames & Synthetic & Yes & 4 K-UHD & YouTube video for HDR + super-resolution \\
\hline

\end{tabular}
\end{center}
\vspace{-14pt}
\label{table:data_set}
\end{table*}

\begin{figure}[t!]
    \centering
    \captionsetup{font=small}
    \includegraphics[width=.93\columnwidth]{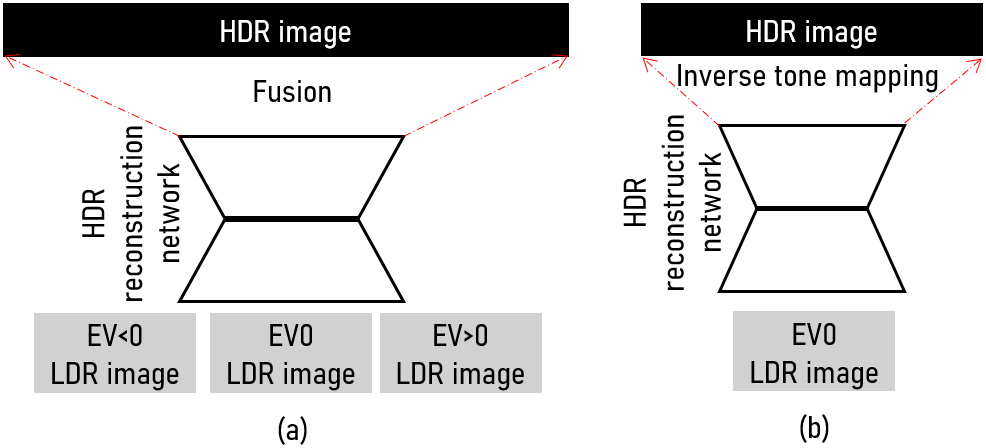}
    \vspace{-8pt}
    \caption{DNN-based HDR imaging approaches. Learning to reconstruct an HDR image (a) by aligning, merging, and fusing multi-exposure LDR images; (b) by using an inverse one-mapping network from single exposure LDR images.}
    \label{fig:dnn_hdr_category}
    \vspace{-14pt}
\end{figure}

Therefore, deep HDR imaging restores scene irradiance using feature learning from LDR images.
The reconstructed HDR images must satisfy three criteria: high contrast ratio, high bit depth, and preserved details. Fig.~\ref{fig:dnn_hdr_category} shows the pipelines used for deep HDR imaging. The first pipeline, shown in Fig.~\ref{fig:dnn_hdr_category}(a) aims to create an HDR content by learning from multi-exposure LDR images with DNNs \cite{kalantari2017deep,yan2019attention}. Specifically, multi-exposure LDR images can be mapped to the irradiance domain using DNNs based on the inverse CRF and exposure time ($\Delta t$) (or exposure value (EV)\cite{chen2020learning}) from Eq. (\ref{eq1}) and Eq. (\ref{eq2}). Accordingly, an HDR image can be reconstructed from DNNs, which are trained to align and merge dynamic information from the LDR images. It is to be noted that alignment is only required for dynamic scenes.
By contrast, reconstructing HDR images from a single-exposure LDR image is much simpler, as shown in Fig.~\ref{fig:dnn_hdr_category}(b), where an inverse tone-mapping network is learned to generate an HDR image reflecting natural illumination. However, as a single-exposure LDR image conveys limited exposure information, learning the HDR visual content from it is an ill-posed problem \cite{banterle2017advanced,chen2020learning}. Therefore, the reconstructed HDR image has a limited dynamic range, which indicates that the visual quality of the HDR images is inferior to that of the multiple-exposure LDR images. As HDR images are usually displayed after tone mapping, most DNN-based methods optimize learning on tone-mapped HDR images, which is more effective than direct computation in the HDR domain \cite{kalantari2017deep,yan2019attention,HDRI_github}. A typical approach for learning deep HDR imaging was proposed by \cite{kalantari2017deep}:  the $\mu$-law, which is commonly used for range compression in audio processing.
In Sec.~\ref{hdr_recon}, we conduct a comprehensive and analytical study on existing deep HDR imaging methods. 

\vspace{-7pt}
\subsection{Datasets for Deep HDR Imaging}
The dataset is important for the success of deep HDR imaging methods. Various benchmark datasets available for deep HDR imaging differ in exposure levels, dataset sizes, image resolution, quality, scene diversity, etc. Some datasets obtain LDR images by simulating the CRF or using a CRF database, whereas others contain real-world LDR images with ground-truth (GT) HDR images. In addition, some datasets have been developed for multiple-exposure fusion (MEF) \cite{xu2020fusiondn,xu2020mef}, video HDR \cite{chen2021hdr,jiang2021hdr}, and HDR imaging with novel sensors \cite{han2020neuromorphic,kniaz2018thermalgan}. Table~\ref{table:data_set} summarizes the commonly used benchmark datasets for deep HDR imaging, including the data size, data type, availability of the GT, spatial resolution, and scene details.

Regarding synthetic HDR datasets, HDR GT images are collected from diverse real-world scenes. For single-exposure HDR image reconstruction, the most common approach is to imitate the LDR formation pipeline using different CRF functions from a set of CRF databases \cite{liu2020single,endoSA2017} or a virtual camera to capture a number of random regions of the scene based on randomly selected camera curves \cite{eilertsen2017hdr,santos2020single, lee2018deep}. In \cite{liu2020single}, to create a real-world single-image HDR dataset, 600 amateurs were instructed to capture scenes with multiple exposures using an LDR camera placed on a steady tripod. The LDR exposure stacks were fused to obtain HDR images using Photomatix \footnote{https://www.hdrsoft.com/}. 

Regarding multi-exposure HDR image reconstruction, the dataset collection process from \cite{kalantari2017deep} is representative. For a static scene, HDR image generation is similar to that in \cite{liu2020single}. For the dynamic scene, the HDR image is aligned with the reference middle-exposure LDR images. Using the reference LDR images, the up- and down-exposed LDR images are captured by asking the subject to move simulating camera motion or no motion. More details are provided in the supplementary material.

\vspace{-10pt}
\subsection{HDR Image Quality Assessment (IQA)}
\label{iqa_sec}
Evaluation of the HDR image quality is also essential for deep HDR imaging. In low-level vision, IQA comprises subjective and objective computation approaches \cite{jia2017blind}. We now introduce several commonly used IQA methods, including statistical measures and HDR-oriented IQA measures.

\vspace{-5pt}
\subsubsection{Difference Measures and Statistical-Oriented Metrics}
\noindent \textbf{Peak Signal-to-Noise Ratio (PSNR).} PSNR is a commonly used IQA to assess the quality of reconstruction of lossy transformations \cite{wu2018deep,endoSA2017}. In HDR imaging, PSNR measures the difference between the GT HDR image and the reconstructed HDR image both before and after tone mapping using the $\mu$-law. 

\vspace{3pt}
\noindent \textbf{Mean Squared Error (MSE).} MSE is the simplest yet most widely used full-reference quality metric, computed by averaging the squared intensity differences of reconstructed HDR and GT HDR image pixels before and after tone mapping using the $\mu$-law. 

\vspace{-5pt}
\subsubsection{Structural Similarity Measures}
Structural similarity (SSIM) measures the quality of an image based on pixel statistics to model luminance (using the mean),
contrast (variance), and structure (cross-correlation) \cite{hanhart2015benchmarking}. In deep HDR imaging, SSIM measures the structural differences between the reconstructed HDR and GT HDR image pixels before and after tone mapping using the $\mu$-law. Other structural similarity measures include the multi-scale SSIM index (MS-SSIM) and MEF-SSIM \cite{ma2015perceptual} for image fusion quality measures.

\vspace{-5pt}
\subsubsection{Tailored HDR Statistical Measures}
In HDR imaging, HDR-VDP-2 \cite{mantiuk2011hdr}, which is an updated version of HDR-VDP \cite{mantiuk2005predicting}, is regarded as an effective metric for measuring the visual quality of HDR images. HDR-VDP-2 compares the GT HDR image and the reconstructed HDR image, and predicts visibility and quality. For a detailed discussion on HDR-VDP-2, readers can refer to \cite{mantiuk2011hdr}. Moreover, HDR-VQM \cite{narwaria2015hdr} was designed to evaluate the quality of HDR videos.

\vspace{-5pt}
\subsubsection{Deep Learning--Based Evaluation Metrics}
Compared with conventional HDR image reconstruction methods, DL-based methods are more capable of recovering the dynamic range of images, even in extreme scenes. This has inspired many attempts to employ DNNs for the IQA of HDR images. 
\cite{jia2017blind,artusi2019efficient,he2018quality,kinoshita2019deep,choudhury2018hdr,ziaei2018efficient}. Among these methods, \cite{choudhury2018hdr,artusi2019efficient,ziaei2018efficient} are full-reference methods that assess the perceptual quality of generated HDR images. By contrast, \cite{he2018quality,jia2017blind} are non-reference methods that measure the reconstructed HDR image quality by quantifying the extracted low- and high-level features from DNNs. 


\vspace{-5pt}
\subsubsection{HDR Imaging Challenge}
New trends in image restoration and enhancement (NTIRE)~\cite{perez2021ntire}, in conjunction with CVPR 2021, started the first HDR imaging challenge. It focused on reconstructing an HDR image from a single (Track 1) or multiple (Track 2) exposure LDR images. The workshop utilized existing HDR datasets, \eg,~\cite{kalantari2017deep}, and newly curated HDR datasets for evaluation. The performance was measured by PSNR using the ground truth.  

\vspace{-8pt}
\section{Supervised HDR Reconstruction}
\label{hdr_recon}

\begin{figure*}[t!]
    \centering
    \captionsetup{font=small}
    \includegraphics[width=.98\textwidth]{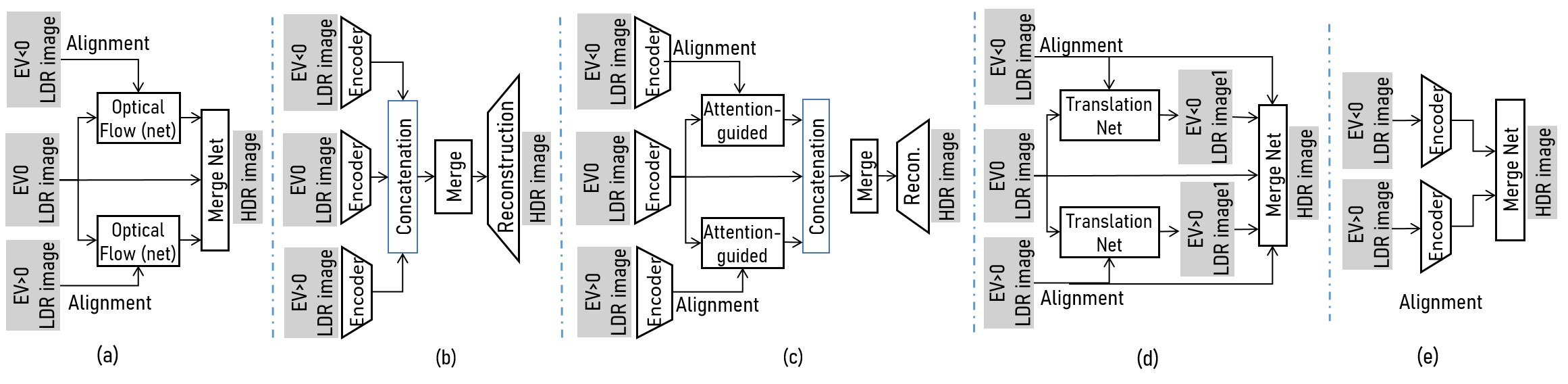}
    \vspace{-13pt}
    \caption{Deep multi-exposure HDR imaging methods. (a) Optical flow-based (net) LDR image alignment. (b) Direct concatenation of features of LDR images to merge them and reconstruct the HDR image. (c) Correlation-guided LDR feature alignment, followed by merging and reconstruction. (d) Image translation-based LDR image alignment. (e) Reconstruction of the HDR image from static LDR image fusion.}
    \label{fig:multiexpo_hdr_category}
    \vspace{-8pt}
\end{figure*}

\subsection{Multi-Exposure HDR Image Reconstruction}
\label{multi-expo}
\noindent \textbf{Insight:} \textit{Reconstructing an HDR image from multiple-exposed LDR images in dynamic scenes follows two steps: (1) learn to align the LDR images; (2) merge them to get an HDR image.
}

The common method of obtaining an HDR image is to capture a series of LDR images at different exposures and merge them to reconstruct an HDR image \cite{choi2020pyramid,kalantari2017deep}. For a dynamic scene, the input LDR images must first be aligned. This step is crucial, yet challenging, for multi-exposure HDR imaging pipelines. The quality of the alignment is usually reflected in the reconstructed HDR images. In other words, a less reasonable alignment often leads to visible artifacts in the pixel regions of low correspondences. We analyze these methods, as shown in Fig.~\ref{fig:multiexpo_hdr_category}.

\vspace{-5pt}
\subsubsection{Optical Flow-based Image Alignment}
A common method of alignment for LDR images is the use of optical flow algorithms or networks. The aligned LDR images are fed to the DNNs to reconstruct an HDR image, as depicted in Fig.~\ref{fig:multiexpo_hdr_category}(a). Given
a sequence of LDR images with low, medium, and high exposures,
Kalantari \etal~\cite{kalantari2017deep} proposed a representative approach aligning low- and high-exposure images to a medium-exposure image, called the reference image, using the classical optical flow algorithm~\cite{liu2009beyond}. The aligned LDR images are taken as inputs to a DNN to reconstruct an HDR image supervised by the GT HDR image. Yan \etal~\cite{yan2019multi} also followed this pipeline. However, to enrich the image information, their method takes a sequence of LDR and HDR images obtained by gamma correction as the inputs at different scales and adopts three sub-networks to obtain the corresponding HDR images. Peng~\etal~\cite{peng2018deep} and Prabhakar \etal~\cite{prabhakar2019fast} argued that classical optical flow algorithms could lead to considerable misalignment errors. They proposed the alignment of low- and high-exposure images with the reference images using SoTA optical flow networks, \eg, FlowNet\cite{dosovitskiy2015flownet}.

\vspace{-8pt}
\subsubsection{Direct Feature Concatenation}
Optical flow algorithms may provide unsatisfactory performance in aligning LDR images with large-scale foreground motion. Therefore, Wu~\etal~\cite{wu2018deep} proposed a representative framework including three encoder networks, a merger network, and a decoder network, as shown in Fig.~\ref{fig:multiexpo_hdr_category}(b). The decoders are trained to encode the LDR images at different exposures into latent feature spaces. Next, the encoded LDR images are fed to a merger to learn the aligned features. Finally, the decoder reconstructs the HDR image. A similar strategy was adopted in~\cite{yan2021towards}; however, in contrast to \cite{wu2018deep}, LDR sub-networks are added to recover three static LDR images corresponding to the HDR image. The feedback, \ie, cycle, reconstruction of LDR images adds more constraints on the forward path, which benefits the restoration of the motion areas in the HDR image. Another end-to-end single processing pipeline was proposed by Chaudhari~\etal~\cite{chaudhari2019merging}; individual encoders are designed to map the raw color filter array (CFA) data of different exposures to learn intermediate LDR features. Metwaly~\etal~\cite{metwaly2020attention} also employed three encoders to extract features of three LDR images at different exposures, similar to \cite{wu2018deep}. The extracted features are concatenated and fed into a merger network. A pivotal contribution of this method is the attention mask, designed to enable the network to focus on the parts of the scene with considerable motion to avoid ghosting effects after the decoder.   
However, Niu \etal~\cite{niu2021hdr} argued that using individual encoders cannot fully exploit multiscale contextual information from features. Therefore, they proposed multiscale LDR encoders to extract visual features at different scales, similar to the strategy proposed by~\cite{yan2019multi}. The extracted features are fused at different scales via residual learning before being fed to a merger network.    


\begin{table*}[t!]
\captionsetup{font=small}
\caption{Deep multi-exposure HDR reconstruction employed by some representative methods. N/A: not available.}
\vspace{-10pt}
\small
\begin{center}
\begin{tabular}{c|c|c|c|c|c|c}
\hline 
Method & Publication &  No. of Exposures & Alignment & Merge & Scene types& Loss functions \\
\hline \hline
Kalantari~\cite{kalantari2017deep} & Siggraph'17 & 3 & Classical optical flow & Image fusion& Dynamic & L2  \\
Prabhakar~\cite{prabhakar2019fast} & ICCP'19 & 3 & Flow+Refine nets & Feature fusion& Dynamic & L2 + MS-SSIM  \\ 
Yan~\cite{yan2019multi} &WACV'19 & 3 & Same as \cite{kalantari2017deep} & \thead{Multiscale \\ image fusion} & Dynamic & L1  \\ 
Wu~\cite{wu2018deep} & ECCV'18 & 3& Direct feature cat. & Feature fusion & Dynamic & L2\\
Yan~\cite{yan2021towards} &Neurocomput. & 3 &Direct feature cat. &  Feature fusion &  Dynamic & L2 + Cycle losses\\
Chaudhari~\cite{chaudhari2019merging} &WACV'20 &$\ge$3 & Feature encoding & Feature fusion & Dynamic & L2 \\
Niu~\cite{niu2021hdr} & TIP'21 & 3 & \thead{Multiscale \\feature cat.}  & \thead{Multiscale \\ feature fusion} & Dynamic &\thead{Two L1 losses\\ + GAN loss} \\
Yan~\cite{yan2019attention} & CVPR'19 & 3 & \thead{Attention-guided\\
feature alignment}& \thead{Dilated dense \\feature merge} & Dynamic &L1  \\
Yan~\cite{yan2020deep} &TIP'20& 3 & Direct feature cat. & \thead{non-local + triple-pass \\
feature correlations} &Dynamic & L2  \\
Choi~\cite{choi2020pyramid} & Sensor'20 & 3 & \thead{Pyramid-attention \\feature alignment} & \thead{Dilated dense \\feature merge} & Dynamic & L1  \\
Pu~\cite{pu2020robust} &ACCV'20 &3 & \thead{Multiscale \\pyramidal
alignment} &  \thead{Residual dense \\ feature merge} & Dynamic & L1 + Gradient \\ 
Rosh~\cite{ks2019deep} & ICIP'19 & 3 & Image translation & \thead{Multi-exposure \\image merge}& Dynamic & L1 + Perceptual \\
\hline
Prabhakar\cite{prabhakar2017deepfuse} & CVPR'17 & 2 & N/A & \thead{Unsupervised\\ feature fusion} & Static & SSIM \\
Ma\cite{ma2019deep} & TIP'19 & $\ge$3 & N/A & \thead{Guided weight \\ map filtering} & Static & FEM-SSIM \\
Xu\cite{xu2020mef} & TIP'20 & 2& N/A & \thead{Self-attention \\ + local detail cat.} & Static & \thead{MSE + GAN \\ + Gradient}\\
\hline

\end{tabular}
\end{center}
\vspace{-12pt}
\label{table:multi_expososure_hdr}
\end{table*}

\vspace{-8pt}
\subsubsection{Correlation-Guided Feature Alignment}
In reality, object motion in a dynamic scene can be diverse. This makes it difficult to suppress ghosting effects in flow-based HDR image reconstruction methods.
Recent DL-based methods have been explored to determine the correlation between the LDR inputs of different exposures and motions. Using the attention mechanism \cite{vaswani2017attention} is a representative approach to these problems, as shown in Fig.~\ref{fig:multiexpo_hdr_category}(c). Yan \etal \cite{yan2019attention} generally followed the pipeline of \cite{wu2018deep}. However, their method applies the attention modules in the encoding phase to achieve better feature alignment. In particular, attention modules help exclude less useful components caused by misalignment and saturation in individual encoders. This is done by feeding under- or over-exposed LDR images (non-reference LDR images) and the reference LDR image to two attention modules. Attention maps are then obtained to refine the feature maps of the non-reference LDR images. Consequently, misaligned and saturated non-reference LDR images are excluded, preventing less useful features from entering the merging network.  

Pu \etal~\cite{pu2020robust} and Choi~\etal \cite{choi2020pyramid} explored the alignment of the features of LDR images using the pyramid modules in the encoding stage. This is conceptually similar to that in \cite{yan2019attention}. In particular, the method in~\cite{pu2020robust} proposes a multiscale pyramidal feature alignment strategy. This is more flexible and robust in handling motion and disparity. The method in~\cite{choi2020pyramid} explores the inter-attention of the self-similarity of pixels in LDR images. A dual excitation block is also designed to recalibrate the features spatially and channel-wise in the merging phase. By contrast, the method proposed by~\cite{yan2020nonlocal} employs non-local modules \cite{vaswani2017attention} in the merging phase instead of the encoding phase. The encoded features are directly concatenated and fed into the merging network.
The method proposed by~\cite{chen2020deep} uses only two LDR images (one under-exposed and one over-exposed) and adopts a homograph network (encoder) to warp the under-exposed images to the overexposed images. Then, an attention module is employed to reduce the misaligned features before being fed to the merging network. In the NTIRE 2021 HDR Challenge~\cite{perez2021ntire}, ADNet~\cite{liu2021adnet} proposes the alignment of LDR images with a pyramid, cascading, and deformable (PCD) module and adaptively fuses them with a spatial attention module. 

\vspace{-5pt}
\subsubsection{Image Translation-Based Alignment}
Although optical flow-based HDR imaging methods can reconstruct ghost-free HDR images in conditions with varied motions and different exposures, these methods tend to generate considerable artifacts. Moreover, the end-to-end feature concatenation-based methods successfully reconstruct HDR images; however, they generate less realistic details in highly saturated regions. Some studies have attempted to translate low- and high-exposure images to the reference LDR image using DNNs under the supervision of the reference image, as depicted in Fig.~\ref{fig:multiexpo_hdr_category}(d). 
The method proposed by Rosh \etal~\cite{ks2019deep} is a representative method in which image translation networks are designed to generate two LDR images resembling the reference LDR image. The three LDR images are then fed to a merging network to reconstruct an HDR image; however, the two stages are not learned in an end-to-end manner. Lee \etal~\cite{lee2020exposure} extended \cite{ks2019deep} and generated accurately aligned LDR images by blending information from multi-exposure images using encoder--decoder network structures. The alignment and merging stages can be trained in an end-to-end manner.

\vspace{-3pt}

\subsubsection{Deep Static Exposure Fusion}

Compared with multi-exposure HDR imaging in dynamic scenes, merging LDR images with static scenes is relatively simple. The main challenge is to find methods that merge LDR images at different exposures and address the artifacts and loss of textural details in the reconstructed HDR images. In the literature, MEF is a method of merging multi-exposure images. The MEF strategy has been broadly studied for image enhancement and HDR imaging. This study reviews and analyzes representative MEF methods targeted at HDR imaging using DNNs. For details on more general image fusion techniques, refer to \cite{kaur2021image}. 

\begin{figure*}[t!]
    \centering
    \captionsetup{font=small}
    \includegraphics[width=.99\textwidth]{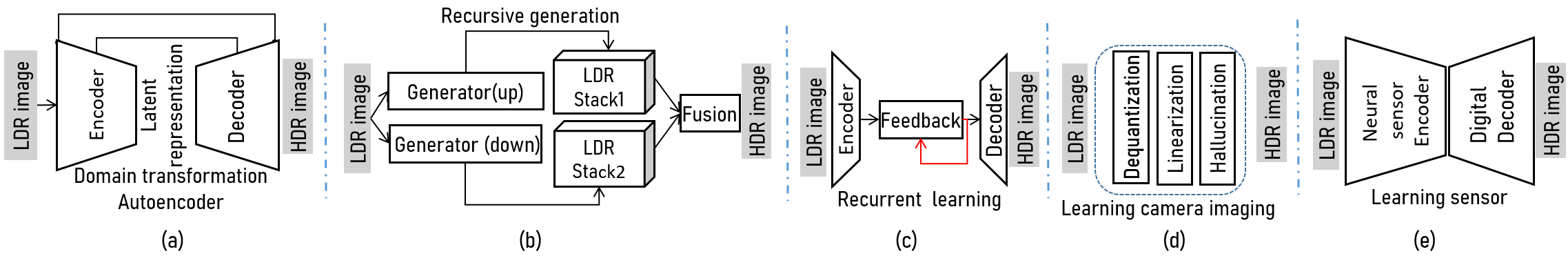}
    \vspace{-10pt}
    \caption{Deep single-exposure HDR imaging methods. (a) Directly learning a domain transformation autoencoder from a single LDR image. (b) Generating bracketed (up- and over-exposed) LDR image stacks and then reconstructing an HDR image. (c) Using a recurrent (efficient) learning network structure. (d) Learning camera-imaging pipeline with DNNs. (e) Learning camera sensing with DNNs (neural sensing encoder).}
    \label{fig:single_hdr_category}
    \vspace{-8pt}
\end{figure*}

 DeepFuse, proposed by Prabhakar \etal~\cite{prabhakar2017deepfuse}, is a representative learning approach for static multi-exposure image fusion. To achieve this fusion, DeepFuse collects a large dataset of multi-exposure image stacks for training and reduces the need for GT images. The unsupervised DL
framework for the MEF uses a no-reference image quality loss function. This baseline has been updated by many recent studies \cite{ma2019deep,xu2020fusiondn,yin2020deep,xu2020mef}. MEF-Net, proposed by Ma \etal~\cite{ma2019deep}, focuses on static LDR image fusion
of arbitrary spatial resolution and exposure levels. It typically combines single-image super-resolution (SISR) and HDR imaging by feeding a low-resolution (LR) version of the input sequence to a neural network for weight map prediction. The predicted weight maps are jointly
upsampled using a guided filter. The final HDR image is calculated by weighted fusion with high-resolution (HR) LDR images. MEF-GAN proposed by Xu~\etal~\cite{xu2020mef} is targeted at multi-exposure image fusion based on GANs \cite{goodfellow2014generative, wang2020deceiving}. As the luminance of the multi-exposure images
varies greatly with the spatial location in LDR images, the self-attention mechanism \cite{yan2020nonlocal} is adopted in the generator to learn the attention-driven and
long-range dependency of the LDR images. FusionDN~\cite{xu2020fusiondn} extends DeepFuse and proposes an unsupervised and unified densely connected network for image fusion tasks, including LDR image fusion. Given two LDR images, FusionDN learns to generate a fused HDR image. This is achieved by a weight block, which can obtain two image-driven weights as the measurement of information preservation in the features of different LDR images. Thus, losses of similarities based on weights can be applied without using GT images.

\vspace{-5pt}

\subsubsection{Potential and Challenges}
From the aforementioned review, one crucial yet meaningful step for multi-exposure HDR imaging is LDR image alignment and merging, as summarized in Table~\ref{table:multi_expososure_hdr}. Aligning LDR images using optical flow is a traditional approach; however, this alignment method is error-prone and less effective with large motions. By contrast, correlation-guided feature alignment is more flexible and effective. Therefore, the SoTA methods mostly explore the correlation of features, \eg,  using the attention mechanism to exclude misaligned features. However, correlation-guided feature alignment is sensitive to over-saturated regions, which often causes the loss of textual details owing to the exclusion of features. Some methods, \eg, \cite{yan2019attention}, employ attention in the encoding phase, and other methods, \eg, \cite{yan2020nonlocal}, learn the correlation in the merging phase. Nonetheless, more theoretical studies are needed to determine the phase that is most appropriate for learning correlations. 

Many challenges remain in this field of research. First, a robust image or feature alignment must be guaranteed. The learning correlation of LDR images is a promising research direction; however, it usually requires considerable computational cost. Second, many labeled datasets are required to train DNNs and make them robust. In multi-exposure HDR imaging, it is relatively expensive to obtain cameras and equipment for scene capture. Therefore, future research needs to investigate data-efficient learning. One promising direction is to explore knowledge transfer~\cite{wang2021knowledge} or semi-supervised learning \cite{van2020survey,wang2022semi}. Third, as DNNs are adopted for learning salient features, inference latency is inevitable in most SoTA methods. However, real-time HDR imaging is preferred for real-life applications. Thus, it is crucial to develop lightweight DL-based frameworks for balancing performance. We further discuss these problems in Sec.~\ref{sec:discussions}. 

\vspace{-8pt}
\subsection{Deep Single-Exposure HDR Imaging Approaches}
\label{single-expo}
\noindent \textbf{Insight:} \textit{Single-exposure HDR image reconstruction aims to recover missing information in the saturated areas of an LDR image and reconstruct an HDR image using a neural network.}

Despite the success of deep HDR image reconstruction by combining a set of LDR images at different exposures, this paradigm still has limitations; it has to handle misalignment with reference images due to scene motion or requires specialized optical systems. Single-exposure HDR image reconstruction avoids these limitations. The distinct advantage of single-exposure HDR reconstruction is that it can work with images captured with a standard camera or even recover the full dynamic range of legacy LDR content \cite{santos2020single}. Consequently, single-exposure HDR reconstruction has attracted attention in the research community.

\vspace{-5pt}
\subsubsection{Direct Learning from a Single LDR Image} 
The most straightforward method is to directly learn from a single LDR image using encoder--decoder network structures, as shown in Fig.~\ref{fig:single_hdr_category}(a).
HDRCNN, proposed by Eilertsen \etal~\cite{eilertsen2017hdr}, is a representative approach for reconstructing an HDR image from a single LDR image. In particular, a hybrid LDR encoder and HDR decoder, which operate in the log domain, are used to reconstruct an HDR image. To improve the generalization of complex saturation scenes, HDR images from existing datasets are used to simulate diverse LDR exposures. 
Santos \etal~\cite{santos2020single} identified the disadvantages of applying the same convolutional filters to well-exposed and saturated pixels in~\cite{eilertsen2017hdr}, causing ambiguity during training and leading to checkerboard and halo artifacts in the HDR images. To solve this problem, their method involves masking the saturated areas, thus reducing the contribution of the features from these areas. Meanwhile, a perceptual loss is also introduced to help reconstruct visually pleasing textures.
By contrast, ExpandNet~\cite{marnerides2018expandnet} takes a different viewpoint; this is a multiscale autoencoder architecture that aims to learn a different level of detail from an LDR image. The learned multilevel features are merged to reconstruct an HDR image. The method in \cite{gharbi2017deep} proposes a DNN-based approach to learn the local and global features for HDR imaging at a high resolution.
These seminal methods have inspired many attempts to improve the diversity of exposure to LDR images. Moriwaki \etal~\cite{moriwaki2018hybrid} found that using the reconstruction loss, \eg, mean square error (MSE) loss, could often cause blur effects and the loss of semantic details in HDR images. To solve these problems, perceptual loss and adversarial loss are added to improve the perceptual quality. The method in \cite{kinoshita2019itm} also proposes a novel cosine similarity loss, normalizing the HDR images and distributing the pixel value of HDR images.

\begin{table*}[t!]
\captionsetup{font=small}
\renewcommand{\tabcolsep}{2pt}
\caption{Deep single-exposure HDR imaging by some representative methods.}
\vspace{-8pt}
\small
\begin{center}
\begin{tabular}{c|c|c|c|c|c|c|c|c}
\hline
Method & Input & CRF & \thead{Image \\ stacking} & \thead{Under-exposed \\ regions} & \thead{Over-exposed \\ regions} & Merge& \thead{Learning \\ highlight} & Loss \\
\hline \hline
 Eilertsen~\cite{eilertsen2017hdr} & \parbox[t]{2 mm}{\multirow{15}{*}{\rotatebox[origin=c]{90}{2D Image}}} & Fixed CRF & \xmark &\checkmark & \xmark & \xmark& \thead{Autoencoder with\\ skip connections} & \thead{L2 (Illuminance +\\ reflection)}  \\
Endo~\cite{prabhakar2019fast} & & CRF database & \xmark & \checkmark & \checkmark& \checkmark & 3D Autoencoder & L1 + L1  \\ 
Marnerides~\cite{marnerides2018expandnet} & & \xmark & \xmark &\checkmark & \checkmark & \checkmark & Multi-branch CNN&L1 + Cosine  \\ 
Santos \cite{santos2020single} & & \xmark & \xmark& \checkmark & \checkmark & \xmark & Feature masking CNN & \thead{L1 + Style + \\ Perceptual} \\
Lee \cite{lee2018recursive} & &\xmark & \checkmark &\checkmark & \checkmark & \checkmark & Recursive learning &L1 + GAN \\
Kim \cite{kim2020end} & &\thead{Linear \\inverse CRF} & \checkmark &\checkmark &\checkmark &\checkmark & Recursive learning &\thead{ Global + local \\ + refine} \\
Khan\cite{khan2019fhdr} && \xmark &\xmark &\checkmark &\checkmark & \xmark & Feedback RNN & L1 + Perceptual \\
Liu \cite{liu2020single} && \thead{Deep \\inverse CRF} & \xmark &\checkmark & \checkmark &\xmark & \thead{Dequantization \\ + Hallucination \\+ Refinement} & Joint losses\\
Zeng \cite{zeng2020learning} && \xmark & \xmark &\checkmark &\checkmark &\xmark & \thead{Image-adaptive \\3D LUT CNN} & \thead{L2 + GAN + \\Regularization} \\
\hline
 Zhang \cite{zhang2017learning} & \parbox[t]{10mm}{\multirow{3}{*}{\rotatebox[origin=c]{90}{\thead{3D \\Panorama}}}}  & \xmark &\xmark &\checkmark &\checkmark &\xmark & \thead{3D panorama \\Autoencoder \\+ Regression} & L1 + Render \\
Yang \cite{yang2018image} && \xmark & \xmark & \checkmark & \checkmark & \xmark & \thead{3D Panorama \\ Reconstruction + Correction}  &  L2 + LDR\\
\hline
Metzler  \cite{metzler2020deep} & \parbox[t]{10 mm}{\multirow{3}{*}{\rotatebox[origin=c]{90}{\thead{Neural\\Sensor}}}}  & \xmark& \xmark & \checkmark &\checkmark &\xmark & \thead{Optical encoder + \\ Electronic decoder} & L2 + Regularization \\
 Martel \cite{martel2020neural}&& Fixed CRF &  \xmark & \checkmark &\checkmark & \xmark&  \thead{Electronic encoder + \\ Digital decoder} & L2 \\

\hline

\end{tabular}
\end{center}
\vspace{-12pt}
\label{table:single_hdr_table}
\end{table*}
\vspace{-3pt}
\subsubsection{Generating Bracketed LDR Image Stacks}
In reality, single-exposure HDR reconstruction suffers from a lack of quality training data and is an ill-posed problem. To this end, Endo \etal~\cite{endoSA2017} proposed a representative framework based on a 2D encoder and 3D decoder with skip connections to generate bracketed LDR image stacks (up-exposed and down-exposed) from an HDR image using the camera response function (CRF) database. As such, an HDR image is reconstructed from the bracketed LDR images, as shown in Fig.~\ref{fig:single_hdr_category}(b). This method produces natural tones without introducing visible noise or the color of the saturated regions.
Motivated by \cite{endoSA2017}, Lee \etal~\cite{lee2018deep} generated LDR stacks with a two-branch network. In particular, three up-exposed and three down-exposed LDR stacks are generated sequentially. Thus, six LDR images are obtained and merged to generate an HDR image. Similar approaches were also proposed by \cite{lee2018recursive,kim2020end}. However, multi-exposure stacks are generated using GANs \cite{goodfellow2014generative,wang2020deceiving}. Cai~\etal~\cite{cai2018learning} collected a dataset containing both up- and down-exposed LDR stacks generated from one LDR image by varying the exposure values (EVs). The dataset is then used to train a neural network to reconstruct an HDR image. 
Instead of generating multiple exposures, An \etal~\cite{an2017single} argued that a single image could contain pixel-varying exposure for a scene. Thus, an image is changed to the Bayer mode and multiple sub-images are generated according to each exposure and merged to reconstruct an HDR image using a neural network.   

\vspace{-5pt}
\subsubsection{Computationally Efficient Learning} Although the HDR image quality can be improved by increasing network depth or adding more losses, it involves considerable computational costs. FHDR, proposed by Khan \etal~\cite{khan2019fhdr}, exploits the power of the feedback mechanism, and proposes a recurrent
neural network (RNN)-based frameworks (see Fig.~\ref{fig:single_hdr_category}(c)). As such, low-level features are guided by high-level features in multiple iterations, leading to better reconstruction with few parameters. In contrast to \cite{khan2019fhdr}, a different perspective of the feedback mechanism was proposed in \cite{yang2018image}; the method involves learning to generate HDR images first. Next, the outputs are used to generate LDR images reciprocally via a correction network. As the correction network can be removed after training, there is no additional inference cost. Zeng \etal~\cite{zeng2020learning} found that, in the camera imaging pipeline, 3-dimensional lookup tables (3D LUTs) are important for manipulating the color and tone of photos. Thus, they proposed learning image-adaptive 3D LUTs together with a small network to reconstruct high-resolution HDR images effectively. The method is shown to produce satisfactory results with fewer computation costs.  A more detailed analysis is provided in Sec.~\ref{sec:discussions}. 

\vspace{-5pt}
\subsubsection{Learning Camera Imaging Pipeline}
The major challenge for HDR image reconstruction from a single image is to restore the missing details in the under- or over-exposed regions in the LDR images. The details are missing because of the quantization and saturation of the camera sensors. Therefore, modeling the imaging pipeline is a pivotal step in effective HDR image reconstruction. SingleHDR, proposed by Liu \etal~\cite{liu2020single}, is a representative framework for modeling the HDR-to-LDR formation process into three sub-steps: dynamic range clipping, non-linear mapping with
a CRF, and quantization, as shown in Fig.~\ref{fig:single_hdr_category}(d). Based on these intuitions, tailored DNN models are built to sequentially learn the three reverse steps. The sub-networks are jointly fine-tuned in an end-to-end manner to reconstruct an HDR image. SingleHDR has been demonstrated to perform favorably against methods generating bracketed LDR image stacks,~\eg,~\cite{lee2018recursive}.

\vspace{-5pt}
\subsubsection{Learning Neural Sensors} 
In a single-exposure HDR imaging pipeline, the biggest challenge is to correctly restore the saturated regions of the LDR images. The aforementioned approaches have addressed this problem by generating LDR images from the CRF database, directly modeling the reverse CRF, or designing efficient networks. However, these methods do not consider the problems within the sensors. Recent methods have attempted to model sensor processing using DNNs, which is a promising direction in HDR imaging, as depicted in Fig.~\ref{fig:single_hdr_category}(e). In particular, the method proposed by Metzler \etal~\cite{metzler2020deep} introduces an optical encoder to encode optical HDR information of the lens. Meanwhile, an electronic decoder network is designed to decode optically encoded information to reconstruct an HDR image. A similar approach was proposed in \cite{alghamdi2019reconfigurable}. A modulation method is designed to learn spatially optical coding information; however, the overall framework is an end-to-end learning method.
Martel \etal~\cite{martel2020neural} introduced a differential neural sensor to optimize per-pixel shutter functions, jointly learned by a neural network in an end-to-end manner. Modeling the exposure function enables the sensor to capture blurred LDR images, which are then used to reconstruct HDR images.

\vspace{-8pt}
\subsubsection{Potential and Challenges}
Based on the aforementioned analysis, various aspects of deep single-exposure HDR imaging methods have been studied. The rapid development of methodologies has resulted in significant performance enhancements. Table~\ref{table:single_hdr_table} summarizes representative studies. Deep single-exposure HDR imaging has several advantages. First, multi-exposure HDR imaging excludes the alignment problem of LDR images and suffers from the ghosting effect to a lesser degree. Second, it is more flexible in applications and simplifies data collection. Third, it is computationally efficient.

There are several challenges in this direction. First, it is difficult to estimate the saturated regions of the LDR images. Although learning optical sensors \cite{alghamdi2019reconfigurable,martel2020neural,metzler2020deep} improves the estimation of the saturated pixels, it requires more complex camera settings and hardware. Future research could simplify the hardware settings and learning frameworks. Moreover, mainstream methods attempt to generate LDR images under diverse illuminance conditions; however, in reality, complex saturated pixels are still difficult to handle using a CRF database or learning CRF. It might be more promising to simultaneously combine the camera imaging pipeline and CRF to reinforce a more robust estimation of the saturated pixels in the LDR images.
Furthermore, most methods resort to pixel-wise loss, \eg, L1 loss, for optimization. Although some methods include adversarial loss and perceptual loss to enhance perceptual quality, these losses sometimes lead to less realistic results. We discuss learning strategies in Sec.~\ref{loss_functions}.

\vspace{-10pt}
\subsection{Deep 3D Panorama HDR imaging}
\label{panaroma-hdr}
\noindent \textbf{Insight:} \textit{Deep 3D Panorama HDR imaging aims to reconstruct an HDR panorama by learning the lighting from a single LDR image.}

\vspace{3pt}
Unlike the general 2D image, a 3D panorama represents the physical world in a 3D wide-range view \cite{song2019neural}. Owing to its wide-range view,  lighting in a scene often varies spatially, making it more challenging to estimate 3D lighting for HDR imaging. Zhang \etal \cite{zhang2017learning} proposed a representative method that aimed to learn inverse tone mapping from an LDR panorama image to reconstruct an HDR panorama image in outdoor scenes. The framework typically follows an encoder--decoder structure that reconstructs HDR images and estimates sun elevation. 
By contrast, Gardner \etal~\cite{gardner2019deep}, Gkitsas \etal~\cite{gkitsas2020deep}, and Song \etal \cite{song2019neural} focused on indoor panorama HDR imaging by learning parametric lighting. In particular, in \cite{gardner2019deep}, lighting is represented as a set of discrete 3D lights with geometric and photometric parameters, which are estimated using DNN learning from a single image on a dataset of environment maps annotated with depth. In \cite{gkitsas2020deep}, in contrast, the lighting incoming from all directions at a 3D location is represented by learning from a single LDR image and a selected 2D pixel. The core idea is to decompose the panorama HDR imaging process into three subtask networks: geometry estimation, LDR completion, and LDR-to-HDR mapping. The networks are trained sequentially and then fine-tuned in an end-to-end manner.

\noindent \textbf{Discussion:} Compared with 2D HDR imaging, there are fewer studies on 3D panorama HDR imaging. 3D panorama HDR imaging is a complex task that requires a robust algorithm that can correctly estimate spatially varying lighting. The biggest difficulty in learning 3D panorama HDR imaging is the lack of qualitative and quantitative real-world labeled datasets. Most methods rely on synthetic datasets and learn a parametric lighting model in multiple steps. Such learning methods are expensive and data-driven. Future research may consider bridging 2D HDR imaging with 3D panorama HDR imaging using knowledge transfer \cite{wang2021knowledge,wang2021evdistill}. As it is relatively easy to collect labeled datasets in the 2D domain and there are many pretrained network models, the methods that facilitate 3D panorama HDR imaging can be explored. 

\vspace{-7pt}
\subsection{Deep Stereo-Based HDR Imaging}
\label{sec:stereo_hdr}
\noindent \textbf{Insight:} \textit{Deep stereo-based HDR imaging aims to reconstruct an HDR image from a pair of multi-exposure stereo LDR images.}

\begin{figure}[t!]

    \centering
    \captionsetup{font=small}
    \includegraphics[width=.90\columnwidth]{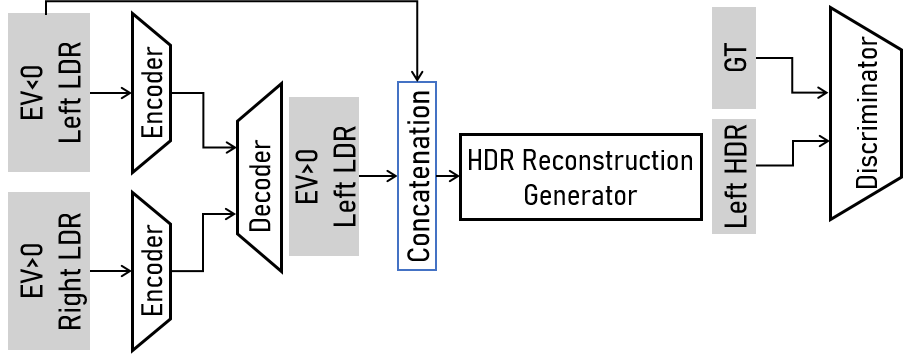}
    \vspace{-5pt}
    \caption{Deep stereo-based HDR imaging method \cite{chen2019new}.}
    \label{fig:stereo_hdr}
    \vspace{-10pt}
\end{figure}

\vspace{3pt}
Stereo-based HDR image reconstruction has been widely studied. Lin \etal \cite{lin2009high} proposed the first stereo-based HDR imaging framework, which includes five steps: CRF estimation, stereo matching, image warping, hole-filling, and HDR image fusion. This baseline framework has been regarded as a basic paradigm for stereo HDR imaging. However, it involves multiple processing steps, and the reconstructed HDR image is highly dependent on the 
performance of each processing step.
Recently, DNNs have been applied to stereo-based HDR imaging. The method proposed by Chen~\etal~\cite{chen2019new} is a representative framework for stereo HDR with an LDR stereo imaging system trained using a GAN, as shown in Fig.~\ref{fig:stereo_hdr}.
This framework consists of two steps: exposure view transfer and image fusion. The left-view (LV) under-exposed image is translated to an LV over-exposed LDR image in the first step. Similarly, the right-view (RV) over-exposed image is translated into an RV under-exposed image. In the image fusion step, the input LV LDR image and the translated LV LDR image are fed to a fusion generator to generate an LV HDR image. The same step is applied to the RV images. Thus, HDR LV and RV image pairs are obtained. By contrast, the method in~\cite{chen2020learning} follows the conventional pipeline of \cite{lin2009high} and replaces it with three learning modules: an exposure calibration network, a hole-filling network, and an HDR fusion network. Consequently, the proposed method better addresses the challenges of distorted brightness and color attributes and detail loss in fused HDR images.  

\vspace{3pt}
\noindent \textbf{Discussion and Challenges:}
Based on the aforementioned analysis, only a few methods exist in this research domain. The SoTA method~\cite{chen2020learning} replaces the conventional pipeline~\cite{lin2009high} with DNNs; however, it complicates learning and increases the computational cost. Moreover, as many modules are used, each module needs to be trained and fine-tuned to reconstruct an HDR image in an end-to-end manner. Importantly, exposure calibration and texture filling are challenging steps.
They often cause a loss of texture details and visual artifacts in HDR images. Therefore, further research is needed to address these problems. Finally, the method in~\cite{chen2019new} simplifies the overall pipeline using GAN; however, it is only focused on one-side (either LV or RV) HDR imaging without considering exposure calibration and image wrapping. It could be beneficial to apply image translation methods, \eg, \cite{zhu2017unpaired} to LV and RV wrapping and use the attention mechanism in the merging phase to better exclude misaligned features.

\vspace{-7pt}
\subsection{Deep Video HDR}
\noindent \textbf{Insight:} \textit{Deep video HDR aims to learn clean HDR image content and temporally consistent frames from an input LDR video.}

\vspace{3pt}
Unlike deep HDR imaging, deep video HDR has received relatively less attention in the community. The main reason is that video HDR must address the temporal consistency. In this situation, the reference image at each frame has different exposures, which poses a challenge for image alignment at each frame.

\vspace{-5pt}
\subsubsection{HDR Videos with Single Alternating Exposure}
A representative method was proposed by Kalantari \etal~\cite{kalantari2019deep}. This method is an updated version of an earlier work~\cite{kalantari2017deep} for HDR video reconstruction from LDR sequences with alternating exposures; it includes two steps. As each frame has different exposures, directly using off-the-shelf optical flow algorithms leads to suboptimal HDR video reconstruction. Therefore, the first step is to align the neighboring frames with the current frame using an optical flow network. The estimated flows are used to warp the neighboring frames and produce a set of aligned images. Finally, a merging network is designed to merge the aligned images and reconstruct the final HDR frames. The method in \cite{xu2019deep} is another representative method that directly reconstructs HDR videos from LDR videos. The LDR-to-HDR mapping is similar to single-image HDR methods~\cite{endoSA2017,ning2018learning}, where gamma correction is applied to convert the LDR video into real scenes. The transformed real scenes are fed into a generator under adversarial learning. A generator with a 3D convolutional autoencoder is designed to address the flickering problem caused by temporal inconsistencies. Moreover, intrinsic loss is introduced to maintain the luminance and color information of the output video. 
The methods proposed by Kim \etal~\cite{kim2020jsi,kim2019deep} are also related to video HDR with super-resolution. However, their methods follow a single-exposure HDR imaging pipeline without considering temporal consistency. This issue is discussed in Sec.~\ref{sec:hdr_sr}.                                      
\vspace{-5pt}
\subsubsection{HDR Videos with Multiple Alternating Exposures}
Despite significant progress in deep multi-exposure HDR imaging, deep video HDR with multiple alternating exposures remains challenging. Chen \etal~\cite{chen2021hdr} and Jiang \etal~\cite{jiang2021hdr} proposed two recent representative video HDR methods. In particular, the method in \cite{chen2021hdr} is based on a coarse-to-fine framework, with multiple alternating exposures as inputs. In the coarse reconstruction phase, the optical flow algorithm in \cite{liu2009beyond} is used to align the neighboring frames with the reference frame blended by a weight estimation network. In the refinement phase, the features of the neighboring frames are aligned to the reference frame using deformable convolutions. Finally, the features are temporally fused to reconstruct an HDR video. By contrast, the method in \cite{jiang2021hdr} uses tri-exposure quad-Bayer sensors. The sensors spatially extend each color Bayer filter to four neighboring pixels. The method employs a feature fusion module that merges features in the feature space to handle motion blur. An attention-based temporal denoising module is proposed to reduce noise and maintain temporal coherence. Meanwhile, a super-resolution module is designed to enhance spatial resolution.   

\vspace{-5pt}
\subsubsection{Discussion and Challenges}
Based on the aforementioned analysis, research on video HDR using DL is still rare in the literature. The main challenge is dealing with temporal inconsistencies. It is common to use optical flow to align the neighboring frames with the reference frame; however, this leads to errors in the saturated regions. Although multiple alternating-exposure video HDR methods aim to reconstruct HDR videos in a coarse-to-fine manner, many modules are needed, which increases the computational cost.  More importantly, a factor that hinders the progress of HDR video reconstruction is the lack of high-quality datasets. Although some synthetic datasets have been built~\cite{chen2021hdr,jiang2021hdr}, there exists a considerable domain gap with real-world data. Future research must consider these problems, and investigate the potential of self-supervised learning and transfer learning for reconstructing HDR videos. Moreover, effective temporal consistency loss functions as studied in many video reconstruction problems are worth exploring.

\vspace{-5pt}
\subsection{Optimization and Loss Functions}
\label{loss_functions}
Optimization and loss functions are important for learning effective HDR imaging. In this section, we analyze the loss optimization strategies used for HDR image reconstruction. Tables~\ref{table:multi_expososure_hdr} and \ref{table:single_hdr_table} summarize the commonly used loss functions. As loss functions are crucial for deep HDR imaging, we conducted experiments to validate the effectiveness of the commonly used loss functions. Owing to space limitations, the details of the loss functions and the experimental results validating the effectiveness of loss functions are discussed in the \textit{supplementary material}.

\begin{figure*}[t!]
    \centering
    \captionsetup{font=small}
    \includegraphics[width=.98\textwidth]{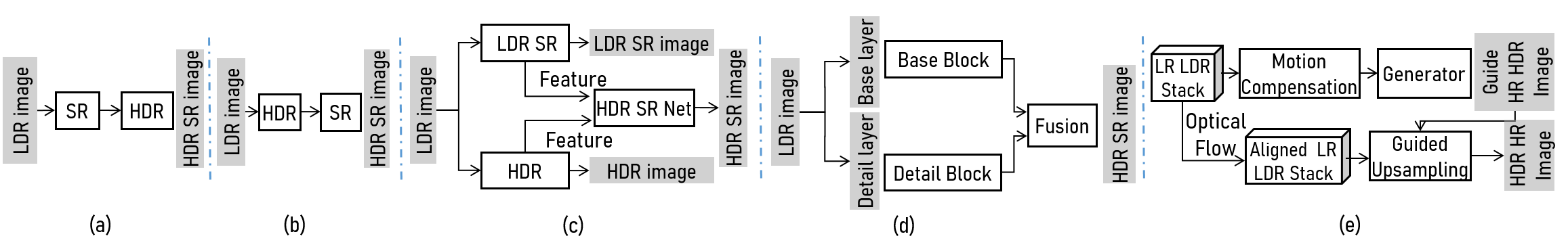}
    \vspace{-13pt}
    \caption{Deep HDR imaging with SR methods. (a) SR is performed first and then the SR HDR image is reconstructed. (b) HDR image is reconstucted first and then SR is performed. (c) LDR SR, LR HDR imaging, and HDR SR are perfomed jointly. (d) LDR images are divided into base and detail layers and fusion of the two branches are learned. (d) SR HDR image is reconstructed from multi-exposure LDR images with optical flow and motion compensation.}
    \label{fig:hdr_sr_category}
    \vspace{-8pt}
\end{figure*}

\subsection{Examining SoTA deep HDR imaging methods}
In the previous sections, we have decomposed the HDR imaging methods based on the number of inputs (Sec.\ref{multi-expo} and Sec.\ref{single-expo}), image modality (Sec.~\ref{sec:stereo_hdr}, and Sec.~\ref{panaroma-hdr}), and loss functions (Sec.~\ref{loss_functions}). Most of the SoTA deep HDR imaging methods can be attributed to combining these components. We have summarized some representative methods and their key strategies in Table.\ref{table:multi_expososure_hdr} and Table~\ref{table:single_hdr_table}. This section presents the quantitative and qualitative results of some representative methods on benchmark datasets. All statistics are derived from the original papers (the source codes are not publicly available) or calculated using official models.

\noindent \textbf{Comparison of the Representative Deep Multi-exposure HDR Methods.} First, we perform a comparison of the representative deep multi-exposure HDR methods based on the taxonomy of Table~\ref{table:multi_expososure_hdr}. \textit{Due to lack of space, the experiments are provided in Sec.4.1 of the supplementary material}.

\noindent \textbf{Comparison of the Representative Deep single image HDR methods.} We then present the experimental results for the representative deep single-image HDR methods, as summarized in Table~\ref{table:single_hdr_table}, on two benchmark datasets. \textit{The experimental results are provided in Sec.4.2 of the supplementary material}.

\vspace{-8pt}
\section{Deep HDR Imaging with other Tasks}
\label{mtl_hdr}
\subsection{Deep HDR Imaging with Super-Resolution (SR)}
\label{sec:hdr_sr}
\noindent \textbf{Insight:} \textit{Deep HDR imaging with SR methods aim to directly reconstruct high-resolution (HR) HDR images from a large amount of low-resolution (LR) LDR image counterparts}.  

\vspace{3pt}
In this section, we provide a perceptive analysis of HDR imaging with SR. A question can be raised: Why do we need to combine HDR with SR? Although ultra-high-definition (UHD) images deliver more realistic visual information, they are less accessible to most users. By contrast, one can easily obtain many LR LDR image resources; therefore, it is important to convert the LR LDR images into HR LDR images. From this perspective, the goal is to solve two visual learning problems: LDR-to-HDR reconstruction and LR-to-HR mapping (SR). Image SR using DL has successfully enhanced performance ~\cite{wang2020eventsr,shocher2018zero}. Similar to HDR imaging, image SR is an ill-posed problem~\cite{wang2020deep,wang2022semi} because multiple HR images exist for an LR image. As the degradation process of LR images is usually complex, the DNN-based methods mostly rely on manually designed degradation kernels~\cite{zhang2018image,shocher2018zero}. Hence, the simultaneous learning of HDR and SR is a complex problem. In LDR images, local variations in contrast and textural details are missing unlike in HDR images. Meanwhile, LR images are degraded by complex degradation kernels with reduced resolution, causing the loss of high-frequency details ~\cite{kim2019deep}. Therefore, it is crucial to recover high-frequency details, contrast, and signal amplitude while enhancing the spatial resolution. 
In general, HDR imaging with SR can be divided into two learning strategies: sequential learning and joint learning. Subsequently, we comprehensively analyze SoTA methods.

\vspace{-5pt}
\subsubsection{Sequential Learning}
Most methods are based on a single image in sequential learning and follow the learning strategy in the image SR. A natural solution of HDR with image SR is sequentially connecting one to the other. However, it is imperative to determine whether the first step should be image SR  (see Fig.~\ref{fig:hdr_sr_category}(a)) or LDR-to-HDR mapping (see Fig.~\ref{fig:hdr_sr_category}(b)). Considering this critical problem, \cite{park2018high, kim2018multi} studied the effect of different connections of modules on the quality of HDR image SR. In particular, in \cite{park2018high}, an image is first decomposed into a luminance component (Y) and chromatic component (UV). Then, two frameworks (HDR-SR and SR-HDR systems) are established to examine learning efficacy. Among many solutions, it has been shown that the best results
can be obtained when processing the luminance component (Y) of the input with a single image and then feeding the enhanced HDR luminance to the single-image SR network trained by
only the luminance component. \cite{kim2018multi} also compared the results of sequential learning with the proposed joint learning framework, which we will discuss in the following section. Their study shows that sequential learning causes an accumulation of errors.




\vspace{-5pt}
\subsubsection{Joint Learning}
Joint learning is the most explored approach to HDR imaging with SR in SoTA methods \cite{zeng2020sr,soh2019joint,kim2018multi,kim2019deep,kim2020jsi}. Although these approaches have different perceptions, they share common features. We examine these methods in this section.

\vspace{3pt}
\noindent \textbf{Single-exposure HDR with SR.} Kim \etal \cite{kim2018multi} proposed a representative framework for joint single-exposure HDR and SR, as shown in Fig.~\ref{fig:hdr_sr_category}(c). The proposed framework includes LDR image SR, LR HDR image reconstruction, and joint HDR SR via feature concatenation. This baseline was improved by \cite{soh2019joint, kim2019deep}, where an LDR image is decomposed into the base layer (illumination) and detail layer (reflectance). Subsequently, two branches are trained based on these layers. Finally, the deep features from the two branches are concatenated to reconstruct an HR HDR image, as shown in Fig.~\ref{fig:hdr_sr_category}(d). A new branch is added in \cite{kim2020jsi}, namely, the image reconstruction module, to the framework of \cite{kim2018multi} to jointly reconstruct the HR HDR image using GAN \cite{goodfellow2014generative}.   

\vspace{3pt}
\noindent \textbf{Multi-exposure HDR with SR.} As single-exposure HDR methods have critical limitations, joint learning of multi-exposure HDR and SR is worth exploring. However, based on this review, few studies have focused on this direction~\cite{deng2021deep,ashwath2020towards}. Ashwath \etal~\cite{ashwath2020towards} proposed a unified framework without joint branches, taking three LR images as inputs, similar to reference-based HDR methods (see Fig.~\ref{fig:hdr_sr_category}(e)). The LR images are first aligned using optical flow and fed to a fusion network to generate an LR feature map. The LR feature map is then fed to a bilateral-guided upsampler to reconstruct an HR HDR image.
By contrast, the method in~\cite{deng2021deep} takes two LR images (under- and over-exposed) with a two-branch learning framework in which each branch aims to learn the under- and over-exposed information recursively. Finally, the information is fused to reconstruct an HR HDR image.

\vspace{-5pt}
\subsection{Deep HDR Imaging with Denoising}
\noindent \textbf{Insight:} \textit{As there exists noise in the under-/over-exposed regions of an LDR image, it is necessary to remove the noise while reconstructing an HDR image.} 

\vspace{-5pt}
\subsubsection{Single-Exposure Noise Removal}
The under-/over-exposed regions of the LDR images may exhibit unexpected noise, affecting the HDR imaging quality. Therefore, SoTA methods attempt to remove noise from the HDR images. Noise2Noise, proposed by~\cite{lehtinen2018noise2noise}, is a representative framework for noise removal in image restoration without using clean data. The method was shown to be effective for HDR imaging of noisy LDR images. 
Godard \etal~\cite{godard2018deep} employed RNNs for burst image denoising. This method handles the noise in HDR imaging better, especially under low-light conditions. Recently, \cite{chen2021hdrunet} proposed a spatially dynamic learning network for HDR imaging with denoising and dequantization. This work won 2nd place in the NITRE2021 single-exposure HDR imaging challenge~\cite{perez2021ntire}.

\vspace{-5pt}
\subsubsection{Multi-Exposure Noise Removal}
Accougalan \etal~\cite{accougalan2020deep} proposed a representative framework for noise removal by merging multi-exposure LDR images in dynamic scenes. Interlacing is a common problem in dual-ISO images. Therefore, a joint deinterlacing and denoising framework is designed to reconstruct a noise-free HDR image. The framework is similar to that of joint HDR and SR~\cite{kim2018multi}, where a network with two branches is used to learn from under-exposed and over-exposed images. The outputs from the two branches are merged to obtain an HDR image. The method in~\cite{accougalan2020hdr} aims to reconstruct clean and noise-free HDR videos from a dual-exposure sensor using a UNet \cite{ronneberger2015u}. As there is a lack of clean GT HDR videos, blur and noise are generated manually. 

\noindent \textbf{Discussion:} From the aforementioned analysis, there are few denoising methods specially designed for single-exposure HDR image reconstruction. This might be because when applying the SoTA denoising frameworks to HDR imaging directly, the reconstructed HDR image tends to lose details~\cite{accougalan2020deep,liu2020single}.
However, more tailored frameworks are required, as the noise at different exposure levels may be different when reconstructing the multi-exposure HDR image. Unlike HDR imaging with SR, few studies have considered joint HDR imaging and image denoising. One reason for this is the difficulty in modeling real-world noise in LDR images. Future research should consider this problem, as real-world LDR images contain considerable noise induced by the sensors in the under-/over-exposed regions. Moreover, similar to HDR imaging with SR, sequential and joint learning effectiveness needs to be studied. 

\vspace{-5pt}
\subsection{Deep HDR Imaging with Deblurring}
\noindent \textbf{Insight:} \textit{In the extreme imaging conditions, \eg, dark scenes, long exposures often cause blur effects in the LDR images. Thus, learning HDR imaging must often consider image deblurring}.

Image deblurring using DNNs has been actively studied in image restoration~\cite{KOH2021103134}. Regarding HDR imaging, the LDR images are often degraded by blur due to the long exposure in extreme conditions. Therefore, some studies have investigated deblurring for HDR imaging.  \cite{aittala2018burst} found that visible light sources and bright highlight regions often appear as elongated
streaks when blurred. To this end, a 
DNN framework was proposed to reconstruct an HDR image from a blurred LDR image. However, it is difficult to restore the saturated regions for single-exposure HDR imaging.  \cite{metzler2020deep} interpreted single-exposure HDR reconstruction as jointly training an optical encoder and electronic
decoder. In the framework, the points are parameterized by the point
spread function (PSF) of the lens. During the inference, a blurred LDR image is fed to the network to reconstruct an HDR image. Similarly, \cite{martel2020neural} introduced a differential neural sensor to optimize per-pixel shutter functions, jointly learned with a DNN in an end-to-end manner. Modeling the exposure function enables the sensor to capture real blurred LDR images used to reconstruct HDR images.

\noindent\textbf{Discussion:} Blur is an important factor to be considered in deep HDR imaging. The methods in \cite{metzler2020deep, martel2020neural} model image sensing with learnable functions to obtain real blurred LDR images. However, these methods are limited to single-exposure HDR imaging. Modeling image sensing could be more challenging for multi-exposure HDR imaging in dynamic scenes. It is possible to combine blind image deblurring methods, \eg,~\cite{Wieschollek_2017_ICCV}, with an HDR reconstruction pipeline. Future research should also consider joint HDR with blind image deblurring.


\vspace{-8pt}
\section{Deep HDR Imaging with Novel Sensors}
\label{hdr_novelsensors}

\begin{figure*}[t!]
    \centering
    \captionsetup{font=small}
    \includegraphics[width=.92\textwidth]{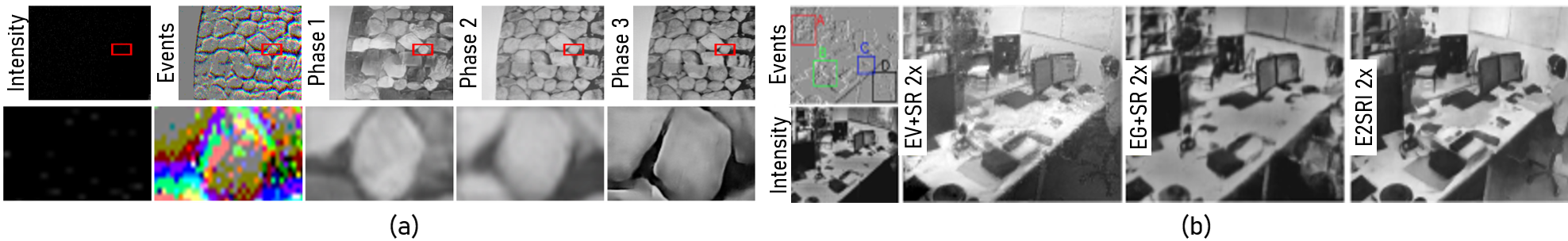}
    \vspace{-7pt}
    \caption{Examples of events in HDR image reconstruction. (a) HDR images from \cite{wang2020eventsr}. From left to right: Dark APS images, embedded LR events, reconstructed LR HDR images, restored LR HDR images, and SR HDR images. (b) HDR images of \cite{mostafavi2021learning} (fourth column), where 2nd and 3rd columns show the reconstructed LR images of \cite{wang2019event} and \cite{rebecq2019high}, respectively.}
    \label{fig:event_hdr_sr_examples}
    \vspace{-10pt}
\end{figure*}

\begin{figure}[t!]
    \centering
    \includegraphics[width=.96\columnwidth]{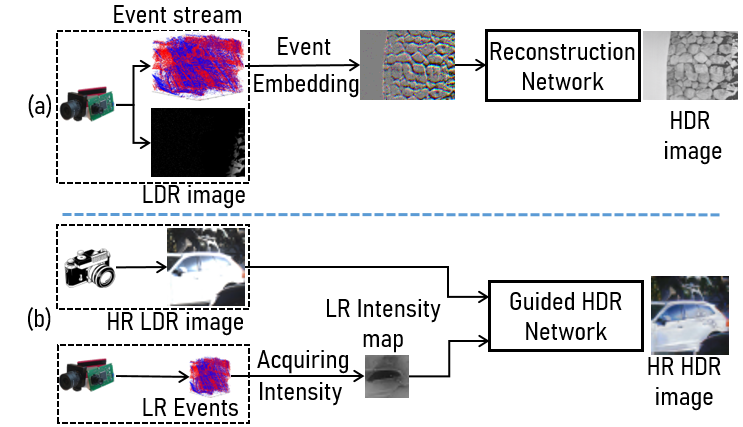}
    \caption{Event-based deep HDR methods. (a) Event-to-HDR image reconstruction. (b) Event-guided HDR image reconstruction~\cite{han2020neuromorphic}. }
    \label{fig:event_hdr_methods}
    \vspace{-10pt}
\end{figure}

\subsection{Deep HDR Imaging with Neuromorphic  Cameras}
\noindent \textbf{Insight:} \textit{Neuromorphic cameras are novel sensors that possess much higher dynamic range than general frame-based cameras, which is promising and advantageous for HDR imaging. }

\vspace{3pt}
Neuromorphic cameras are bio-inspired sensors with a paradigm shift in the manner in which visual information is acquired. Event cameras sample light based on scene dynamics rather than a fixed frame rate, which is unrelated to the viewed scenes. Neuromorphic cameras
are different in principle from conventional cameras~\cite{Gallego2020EventbasedVA}. Unlike conventional cameras that capture intensity values at a fixed frame rate, event cameras sense changes in intensity asynchronously at the time they occur. Neuromorphic cameras, such as DAVIS240C~ \cite{brandli2014240}, have clear advantages over conventional cameras. Specifically, they have a very high HDR ($ 140 dB $ vs. $ 60 dB $).
Active pixel sensor (APS) frames (a type of frame-based intensity sensor) are less invisible in low-light conditions; however, event cameras capture the intensity changes and output convincing visual details. These distinct advantages of event cameras make them promising for robotics, autonomous driving, and wearable applications in scenarios challenging for standard cameras, such as HDR and high-speed motion~\cite{Gallego2020EventbasedVA}. 

\vspace{-5pt}
\subsubsection{Reconstruction-Based HDR Imaging}
As shown in Fig.~\ref{fig:event_hdr_methods}(a), APS frames are less visible when most pixels are under-exposed, but events reflect the scene details with clear edges of objects. To fully explore these visual details, Wang \etal~\cite{wang2019event,mostafavi2021learning} proposed a representative framework for reconstructing HDR intensity images and videos from event data using a conditional generative adversarial network (cGAN). 
Concurrently, Rebecq \etal \cite{rebecq2019high} employed an RNN to reconstruct temporally consistent HDR-intensity videos. Inspired by \cite{wang2019event}, Zhang \etal \cite{zhang2020learning}, and Zhe \etal~\cite{zhe2019dark}, 
further investigated the reconstruction of HDR images from event data under low-light (night) scenes. Some studies further extended the approach of \cite{rebecq2019high} by either minimizing network parameters \cite{scheerlinck2020fast}, improving the generalization of simulated event data \cite{stoffregen2020reducing}, or adding optical flow \cite{zhu2018ev} to impose photometric consistency \cite{paredes2020back} for HDR video reconstruction. Although these methods are effective, the reconstructed HDR results are in LR as event cameras, such as \cite{brandli2014240}, are in LR. To overcome these challenges, Mostafavi \etal~\cite{mostafavi2020learning} and Wang \etal~\cite{wang2020eventsr,wang2021joint} proposed the reconstruction of HR HDR intensity images from LR event data in a supervised and unsupervised learning manner. Visual examples of super-resolved HDR images are shown in Fig.~\ref{fig:event_hdr_sr_examples}. Moreover, \cite{zhang2021event} leveraged the HDR property of the event camera and proposed an event-based synthetic aperture imaging method by combining spiking neural networks (SNNs) and CNNs. The proposed method
collects confident light information from occluded
targets even under extreme lighting conditions, making the
reconstruction of scenes feasible for HDR. Inspired by the HDR image reconstruction methods \cite{wang2020eventsr,wang2019event}, Wang \etal \cite{wang2021evdistill,wang2021dual} proposed to jointly reconstruct HDR images and learn end-tasks under either under-exposed or over-exposed conditions via knowledge distillation \cite{wang2021knowledge}. Moreover, \cite{zou2021learning} proposed a recurrent neural network (RNN) for reconstructing high-speed HDR videos from events.

\noindent \textbf{Potential and challenges:} Event cameras show better HDR capability than frame-based cameras, enabling the aforementioned methods to reconstruct HDR images from events. The reconstructed HDR images bridge the event and intensity domains and can be potentially applied to other vision and graphic tasks. However, some challenges remain. First, reconstructing HDR images using DNNs induces considerable inference latency ~\cite{rebecq2019high,wang2020event}. Second, as events predominantly reflect the edge information of scenes and suffer from noise due to abrupt intensity changes in extreme lighting conditions, the reconstructed HDR images are degraded by noise and loss of textural details. In other words, crucial visual information may be missing during the reconstruction stage.  Finally, the aforementioned methods, except for \cite{wang2021evdistill,wang2021dual}, only consider HDR image reconstruction, with less consideration of its application to end tasks.
Future research should consider jointly learning HDR reconstruction and end tasks to improve efficiency.

\vspace{-5pt}
\subsubsection{Neuromorphic camera-Enhanced HDR Imaging}
The other line of research aims to directly leverage events to guide the reconstruction of HDR images from LDR image content, as illustrated in Fig.~\ref{fig:event_hdr_methods}(b). 
Han \etal~\cite{han2020neuromorphic} proposed a representative framework for single-exposure HDR reconstruction by adding the intensity map information of an event camera. In particular, a hybrid camera system is set up to jointly take a single HR LDR
image and an LR intensity map generated from the events. The proposed HDR reconstruction framework addresses the challenges of gaps in resolution, dynamic range, and color representation between two types of sensors and images. 
Moreover, Wang \etal~\cite{wang2020event} proposed a unified framework by simultaneously denoising, deblurring, and super-resolving the LDR APS images to reconstruct clean HR HDR images. The idea is to integrate the sparse events into a sparsity network, where the LDR APS images and events are jointly learned to reconstruct HDR images. \cite{chen2021indoor} proposed a learning framework to estimate the ambient light using an event camera based on its advantage of obtaining HDR images. 

\vspace{3pt}
\noindent \textbf{Discussion:} Based on our review, only three event-guided deep HDR imaging methods have been proposed in the literature. Han \etal~\cite{han2020neuromorphic} focused on single-exposure HDR image reconstruction guided by an event-generated intensity map.
There are three critical problems associated with this method. First, the intensity map is obtained by running off-the-shelf intensity reconstruction networks, such as \cite{rebecq2019high}. Therefore, there is no direct optimization of HDR image reconstruction and intensity map acquisition. Second, it is limited to single-exposure HDR image reconstruction; 
this limits its ability to handle diverse LDR scenes.  Finally, the HR image sensor and LR event sensor are not perfectly aligned, distorting the reconstructed HDR images. Future research directions should consider end-to-end HDR image reconstruction by jointly learning intensity reconstruction and LDR-to-HDR image mapping. Importantly,  multi-exposure HDR reconstruction approaches, enhanced by multiple event stacks, should be considered to process more complex saturated regions of the LDR images. Moreover, a cross-modality HDR imaging method is explored. The corresponding research question is whether it is possible to directly fuse image and event data and learn a unified HDR imaging framework without relying on intensity reconstruction.

\vspace{-5pt}
\subsection{Deep HDR Imaging with Thermal Sensors}
\noindent \textbf{Insight:} \textit{Thermal or infrared (IR) cameras can highlight thermal objects of the scene and output contour information in a night-time environment, enabling HDR imaging via intensity reconstruction or feature fusion with LDR images.}

\vspace{-5pt}
\subsubsection{Thermal to HDR Image Generation}
IR cameras possess distinct HDR capabilities, especially under low-light conditions. Accordingly, translating an IR image into an HDR color image could be a good solution for enhancing scene perception at night \cite{liu2020deep}. However, as IR images are grayscale and only reflect the object contour, transforming IR images into HDR color images is challenging. Therefore, recent DNN-based methods \cite{kuang2020thermal,goswami2021novel,liu2020deep} aim to learn transformation mapping from the IR image domain to the HDR color image domain. To better preserve the texture information from IR images, many methods have been developed using GANs \cite{goodfellow2014generative,wang2020deceiving}. The method, proposed by Kuang \etal~\cite{kuang2020thermal}, is a representative framework in which a coarse-to-fine generator was used to preserve the textural details of IR images and generate color HDR images in a supervised manner, as shown in Fig.~\ref{fig:thermal_hdr_methods}(a). By contrast, IR-GVI in~\cite{liu2020deep} assumes that no GT color HDR images are available. IR-GVI thus proposes an unsupervised method to map IR images to HDR grayscale images and colorize these images, as shown in Fig.~\ref{fig:thermal_hdr_methods}(b). The method in 
\cite{goswami2021novel} considers GAN-based approaches to be complex and proposes a deep HDR color image reconstruction framework based on the registration with optical image pairs. This method reconstructs high-quality HDR color images and preserves thermal profiles even at night.

\vspace{-5pt}
\subsubsection{Thermal and Image Fusion}
\begin{figure}[t!]
    \centering
    \captionsetup{font=small}
    \includegraphics[width=.94\columnwidth]{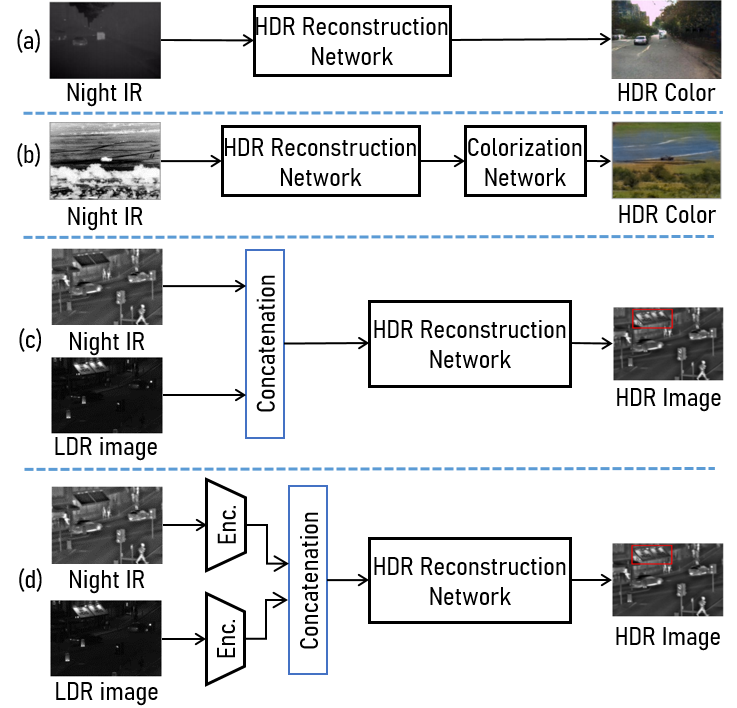}
    \caption{Thermal camera-based HDR methods. (a) Supervised color HDR reconstruction method (b) Unsupervised color HDR reconstruction methods with two steps \cite{liu2020deep}. (c) IR and LDR image concatenation-based HDR reconstruction methods (d) IR and LDR feature concatenation-based HDR reconstruction method.}
    \label{fig:thermal_hdr_methods}
    \vspace{-15pt}
\end{figure}

The other line of research explores fusing IR and LDR images for color HDR image reconstruction~\cite{ma2019infrared}, similar to the multi-exposure HDR imaging pipeline~\cite{eilertsen2017hdr,yan2019attention}. The intuition is that IR images possess grayscale HDR content, but in a different domain than LDR images. Therefore, the challenge is to merge heterogeneous image content or features to reconstruct HDR images. Based on our review, two approaches are commonly used. The first approach involves stacking IR and LDR images, as shown in Fig.~\ref{fig:thermal_hdr_methods}(c) before feeding the results to the reconstruction network~\cite{ma2020infrared,ma2019fusiongan,li2018infrared} using GAN. The second strategy is based on feature fusion~\cite{li2018infrared,li2018densefuse}, as shown in Fig.~\ref{fig:thermal_hdr_methods}(d). Notably, the method in \cite{li2018infrared} employs both image content fusion and feature fusion. The image content is first decomposed and then fused to be fed into a network. Feature fusion based on the VGG network \cite{simonyan2014very} is applied to fuse the image features.

\vspace{-4pt}
\subsection{Deep Modulo Camera-based HDR Imaging}
Conventional cameras fail to capture the over-exposed or under-exposed regions in a scene, which limits the dynamic range of the output. Therefore, HDR reconstruction may lead to considerable artifacts.
The modulo camera concept was first proposed by~\cite{zhao2015unbounded}, aiming to mitigate the limited dynamic range of conventional cameras using a novel, practical sensor. The sensor resets pixels to zero whenever the maximal value is reached in the exposure time to avoid saturation. 
In \cite{jagatap2020high}, HDR image reconstruction using neural priors was proposed. The important feature of this method is that it requires no training data for the dynamic range enhancement of low-light images. The method proposed by Zhou \etal \cite{zhou2020unmodnet} reformulates the unwrapping of a modulo image into a series of binary labeling problems and proposes a DL-based framework to estimate the binary rollover mask of the modulo images iteratively. To date, deep HDR imaging methods using modulo cameras are still limited. Future endeavors could be directed toward this direction.

\vspace{-5pt}
\subsection{Discussion and Challenges}
Generating HDR images from IR images using GANs is the most common approach; however,  the results are usually worse than those based on fusion methods. This is because some visual information might be lost in the generation process, or IR images possess limited visual details; that is, contours of objects are degraded by noise in the LDR conditions. Moreover, recovering color information remains challenging because it is an ill-posed problem. Importantly, generating HDR images leads to considerable inference latency in high-level tasks, such as object detection. 

\begin{figure}[t!]
    \centering
    \includegraphics[width=.75\columnwidth]{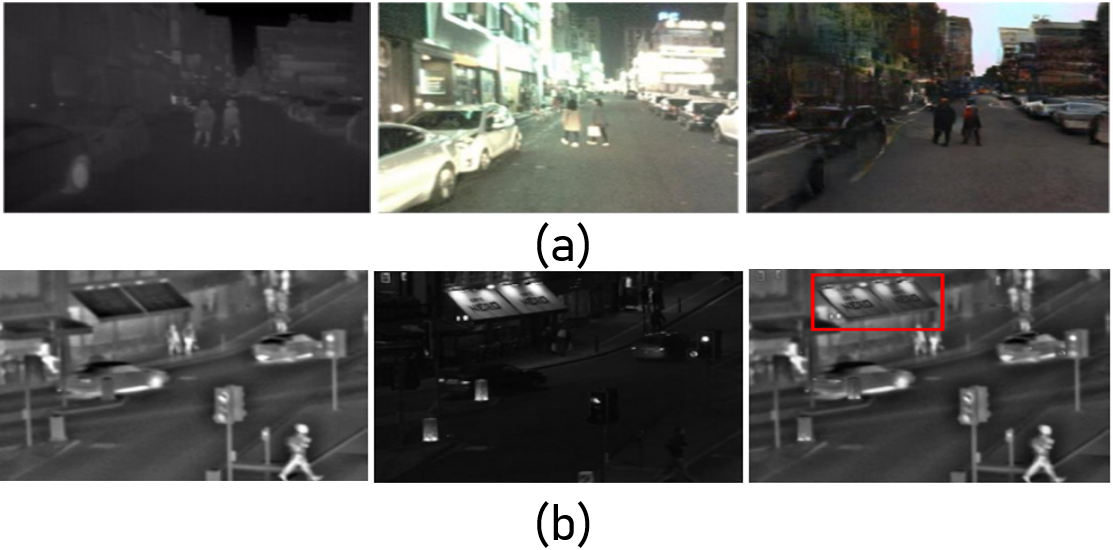}
    \vspace{-8pt}
    \caption{Visual results of IR-based methods. (a) HDR content for image reconstruction \cite{kuang2020thermal}. (b) HDR results from IR and LDR image fusion \cite{li2018densefuse}. From left to right: IR, LDR, and HDR results.}
    \label{fig:thermal_hdr_examples}
    \vspace{-10pt}
\end{figure}

IR cameras show distinct HDR capabilities and are promising for HDR reconstruction, as shown in Fig.~\ref{fig:thermal_hdr_examples}. However, some challenges remain. First, because IR cameras output only grayscale images, most fusion-based methods can only reconstruct grayscale HDR intensity images. Although DenseFuse~\cite{li2018densefuse} can reconstruct color images, color information is not naturally encoded and recovered. Most fusion algorithms are based on GANs and require many IR image training datasets; however, high-quality datasets remain scarce. Our review identifies that most fusion methods concatenate two images or image patch inputs to DNNs. By contrast, fusing features is another approach, as in~\cite{li2018densefuse}; however, this approach treats all features equally, which may not be reasonable in low-light conditions as some pixels in the images are saturated. Therefore, it would be better to filter pixels by considering the degree of exposure in LDR regions.


\vspace{-8pt}
\section{HDR with Novel Learning Strategies}
\label{learning_stratey}

\subsection{HDR Imaging with Unsupervised Learning}
\noindent \textbf{Insight:} \textit{Unsupervised learning reduces the need for GT data and is promising for LDR image fusion and HDR image reconstruction.}

\vspace{3pt}
\noindent\textbf{Unsupervised MEF.} MEF is a type of unsupervised learning that has been actively studied in static LDR image fusion. The MEF aims to fuse LDR images with varying exposures into a single HDR image. DeepFuse \cite{prabhakar2017deepfuse} is a representative unsupervised method that contains three types of layers: feature extraction, fusion, and reconstruction layers. The framework is learned without a reference image using an objective function based on the SSIM image quality measure. The objective function maximizes the structural consistency between the fused image and each input image. \cite{yang2020ganfuse} and \cite{huang2020learning} proposed GAN-based unsupervised frameworks, inspired by CycleGAN~\cite{zhu2017unpaired}, to learn LDR image fusion. \cite{xu2020u2fusion} and \cite{jung2020unsupervised} 
explored the correspondence of source LDR images. As such, the similarity between the fused output and the source images is adaptively preserved.

\vspace{3pt}
\noindent \textbf{Unsupervised LDR-to-HDR Mapping.}
HDR imaging without GT data is challenging, especially for single-exposure HDR image reconstruction. Based on our review, there are no unsupervised methods for single-exposure HDR image reconstruction. For multi-exposure HDR imaging, Li \etal~\cite{li2021uphdr} proposed a representative method called UPHDR-GAN to relax the constraint of paired data. The framework is built on GAN, where the generator takes three LDR images and aims to reconstruct an HDR image. The discriminator distinguishes the tone-mapped image from an unpaired real HDR image. Additionally, perceptual loss is used to preserve the semantic information.

\vspace{3pt}
\noindent \textbf{Discussion:} In our review, most unsupervised methods are targeted at FEM-HDR, and direct LDR-to-HDR mapping methods are rare. Although it is possible to reconstruct an HDR image reconstruction from multi-exposure LDR images without using paired data~\cite{li2021uphdr}, the reconstruction quality is worse than that of supervised methods. Moreover, there is no unsupervised method for single-exposure HDR imaging. Future research may explore more efficient learning in this direction. It could be beneficial to apply contrastive learning \cite{jaiswal2021survey} and domain adaptation \cite{long2016unsupervised} to unsupervised HDR imaging.

\vspace{-5pt}
\subsection{HDR Imaging with GAN}
\noindent \textbf{Insight:} \textit{Using GAN for HDR imaging can help to estimate the true data distribution of light intensity and local contrast.}.

\vspace{3pt}
The key problem in HDR imaging is learning the true data distribution of the light intensity and local contrast. Using L1 or L2 losses to minimize the difference between the HDR image and GT cannot effectively address the problem. By contrast,  
GANs can learn the true data distribution~\cite{goodfellow2014generative,wang2020deceiving,wang2021psat}. Recently, the use of GANs for HDR imaging has attracted attention in the community. In the previous sections, we briefly mentioned the use of GANs for HDR imaging. In this section, we explicitly analyze the technical details of GANs for HDR imaging. We assume that the readers are familiar with the basics of GANs and skip the definitions and formulations. For an introduction to GANs, refer to \cite{goodfellow2014generative}. 

GANs can be applied to HDR imaging as follows: the HDR reconstruction network can be regarded as the generator; then, a discriminator is defined to discern whether it is the generated image or GT. \cite{patel2017generative, ning2018learning} first proposed to add an adversarial loss \cite{goodfellow2014generative} to help train the UNet-based generator \cite{ronneberger2015u} for learning an inverse tone mapping. \cite{rana2019deep} then adopted cGAN \cite{mirza2014conditional} to learn HR and high-quality tone mapping. Considering that the problem of single-exposure HDR imaging is ill-posed, \cite{lee2020learning} proposed generating multi-exposure stacks using cGAN. A cycle consistency loss \cite{zhu2017unpaired} was also used to preserve the exposure level. \cite{kim2020jsi} proposed a GAN-based method for both HDR and SR, where RaGAN \cite{jolicoeur2018relativistic} was adopted as a basic adversarial loss for stable training. All of these methods focus on single-exposure HDR imaging. By contrast, \cite{xu2020mef,niu2021hdr} applied GANs to multi-exposure HDR imaging. Specifically, the method in \cite{xu2020mef} takes under- and over-exposed images and feeds them into two independent feature learning branches, which are then concatenated to reconstruct an HDR image.  A discriminator is used to distinguish whether the fused image was from the GT. The method in \cite{niu2021hdr} uses three LDR images, similar to the reference-based HDR methods, and feeds them to a multiscale LDR image encoder. The extracted features are gradually aggregated to larger scales and finally concatenated to reconstruct an HDR image. A hypersphere adversarial loss~\cite{park2019sphere} was used to train the generator based on the PatchGAN discriminator~\cite{isola2017image}.

\vspace{3pt}
\noindent \textbf{Discussion:}
Although GAN-based HDR methods have shown promising results, some challenges remain. The first challenge is the stability of the training, especially when the LDR data distribution is diverse. Although these methods focus on inverse tone mapping--based (single-exposure) HDR image reconstruction using GAN, they are limited under more complex LDR conditions. It has been shown to achieve multi-exposure HDR image reconstruction with GANs~\cite{niu2021hdr,xu2020mef}; however, only a single discriminator is used to distinguish the entire generated HDR image content from the GT. Indeed, an additional discriminator can distinguish the local regions of the LDR images from the HDR images. Moreover, it is also feasible to deploy a discriminator to facilitate learning using the learned feature representations \cite{wang2020transformation}. 

\vspace{-5pt}
\subsection{HDR Imaging with Transfer Learning}
\noindent \textbf{Insight:} \textit{Transfer learning improves the learning of HDR by transferring knowledge from a related learned task}.

\vspace{3pt}
Recent advancements in DL have shown the possibility of transferring information from a learned task to a target task \cite{wang2021knowledge}. This DL strategy has also been studied for HDR imaging. \cite{alghamdi2021transfer} leveraged transfer learning to overcome the lack of sufficiently large HDR datasets. This demonstrates that transferring knowledge from image classification on ImageNet considerably improves single-exposure HDR reconstruction. \cite{kumar2017no} showed the possibility of transferring the learned information from a related larger database to a smaller one. In particular, AlexNet \cite{krizhevsky2012imagenet} is used to extract features that are adapted to the images of a small database by reducing the dimensionality of the feature vector.

\vspace{3pt}
\noindent \textbf{Discussion and Potential:} From the aforementioned analysis, it is apparent that transfer learning has been less explored for HDR imaging. \cite{alghamdi2021transfer,kumar2017no} transferred the learned features from the image classification task and adapted them to the learning from limited datasets. Recent developments in transfer learning methodologies have been shown to be effective for learning more robust networks and model compression~\cite{wang2021knowledge}. One intuitive research direction is to fully utilize the trained models using labeled HDR datasets to learn a robust student model on an unlabeled dataset. Moreover, it is also meaningful to aggregate the knowledge from the networks learned on multiple datasets to learn a unified model to address more complex LDR image content.
Additionally, the visual information of GT HDR images is less exploited in existing methods; however, it is feasible to transfer the feature information from the HDR image to the reconstruction network learning LDR image content. Finally, in multi-exposure HDR imaging, the multi-exposed images are only aligned to the reference image; nonetheless, the visual information of the non-reference image was less excavated. It might be a good solution to build a mutual learning framework~\cite{zhang2018deep} among student peers to learn more effective knowledge. 

\vspace{-8pt}
\subsection{HDR Imaging with Meta-Learning}
\noindent \textbf{Insight:} \textit{Meta-learning enables quick learning of nonlinear mapping in HDR imaging given a few specific LDR examples.}

\vspace{3pt}
In HDR image reconstruction, there exists a nonlinear mapping between the radiance in a scene and the recorded pixel values of an LDR image~\cite{kim2019deep}. The SoTA DL methods, \eg, \cite{liu2020single,eilertsen2017hdr}, assume that there exists one consistent nonlinear mapping in single-task learning for all possible scenes. However, every scene has unique nonlinear mapping. Thus, \cite{pan2021metahdr} proposed a meta-learning framework in which the ground of meta-parameters is set to learn the common patterns in the nonlinear mapping. It is reasonable to learn nonlinear mapping to better adapt to a specific HDR image reconstruction task. 

\vspace{3pt}
\noindent \textbf{Potential and Challenges.} 
From the aforementioned analysis, it is obvious that the potential of meta-learning remains unexplored. Although \cite{pan2021metahdr} proposed a framework using the popular meta-learning algorithm MAML~\cite{finn2017model}, the HDR images still suffer from artifacts caused by the less effective loss functions that handle color saturation problems. Future research should explore better meta-learning algorithms to capture nonlinear mapping. The method in \cite{soh2020meta} proposes a meta-transfer learning framework to solve the SISR problem, where the DNNs are first trained using a large-scale dataset. Then, meta-transfer learner is built into the internal information within a single image. From \cite{soh2020meta}, it is possible to combine transfer learning with meta-learning to learn a more robust nonlinear mapping for HDR imaging. Moreover, it is worth exploring a few-shot meta-transfer learning framework \cite{sun2019mtl} to adapt to the HDR image reconstruction network.

\vspace{-5pt}
\subsection{HDR Imaging with the Attention Mechanism}
\noindent \textbf{Insight:} \textit{Attention enables the HDR reconstruction network to focus on the most relevant regions of LDR images. }

\vspace{3pt}
The attention mechanism has been shown to enhance the performance of many computer vision problems. Readers can refer to the seminal work~\cite{vaswani2017attention} for more details. The attention mechanism for HDR imaging has also been actively studied. 
\cite{li2019hdrnet} applied the channel-wise attention to capture the feature interdependences between channels. The attention module is built using global average pooling to obtain the global distribution. \cite{yan2019attention} proposed learnable attention modules to guide the merging process of multiple exposures. In particular, attention modules generate soft attention maps to evaluate the importance of different image regions in reconstructing an HDR image. The feature merge using attention guidance was shown to be effective in obtaining ghost-free HDR results. The method of attention guidance for the multi-exposure merge was further enriched by \cite{yan2020nonlocal}, which exploited the nonlocal correlation in the inputs between the encoder and decoder of the UNet structure \cite{ronneberger2015u}. The non-local block helps build dependency between different exposures, thus removing the ghosting artifacts for HDR imaging. Interestingly, to better align the multi-exposed LDR images, \cite{deng2020multi} proposed a multiscale contextual attention module to obtain multiscale attention feature maps for alignment. By contrast, \cite{PAN2020147} focused on static multi-exposure HDR image reconstruction and proposed the use of channel-wise attention, similar to \cite{li2019hdrnet}, to learn the channel-wise statistics for the under-/over-exposed regions in the LDR images. 

\vspace{3pt}
\noindent \textbf{Discussion:} From the aforementioned analysis, we identified that the attention mechanism is effective for multi-exposure mergers regardless of dynamic or static scenes. Using the attention mechanism leads to fewer ghost artifacts than when using an optical flow. However, as pointed out by~\cite{PAN2020147}, using attention mechanisms still has limitations for LDR images with fewer sharp edges and textures.
Increasing the effectiveness of attention mechanisms requires more parameters for the reconstruction networks, thus increasing the computational cost. Therefore, lightweight attention networks that balance the performance and computation costs should be considered. 

\vspace{-5pt}
\subsection{HDR Imaging with Deep Reinforcement Learning}
\noindent \textbf{Insight}: \textit{Deep reinforcement learning can effectively learn multiple local exposure operations based on policy gradient}.

\vspace{3pt}
Deep reinforcement learning is widely used in many visual learning tasks. HDR imaging has been applied to learn multiple local exposure operations~\cite{yu2018deepexposure}. To apply the policy gradient algorithm, the LDR image is decomposed into multiple sub-images. For different sub-images, different exposures are used according to the policy network. To simplify the framework, adversarial learning is adopted, where the discriminator is regarded as a value function. This simple yet novel reinforced adversarial learning approach shows superior visual results for HDR image reconstruction.

\vspace{3pt}
\noindent \textbf{Discussion:} Currently, there is only one method using deep reinforcement learning to exploit multiple local exposure operations in a single image for HDR image reconstruction. Although the learning strategy is different, this is similar to generating multi-exposure LDR image stacks from a single LDR image. Future research should consider MEF methods using reinforcement learning. Furthermore, in \cite{yu2018deepexposure}, sub-images are obtained using image segmentation. However, this method only considers class-level and non-exposure-level regions. Therefore, it is beneficial to segment images based on exposure information to better learn over-/under-exposed regions. 

\vspace{-10pt}
\section{Applications}
\label{applications}
HDR imaging can generate images with a greater range of illuminance than general LDR images, resulting in a better visual representation of detail in highlights or shadows~\cite{wang2018traffic}. Thus, it is of great interest to apply HDR imaging to address some graphics/vision problems in real-life applications.
In this section, we review SoTA research on the applications of imaging techniques. HDR imaging techniques have been applied to computer vision tasks, such as object recognition, scene segmentation, depth estimation, and visual odometry. In computer graphics, it is mainly used for estimating lighting and illumination.

\vspace{3pt}
\noindent \textbf{Scene Recognition and Detection.}
HDR images naturally reflect light patterns and textural details in dark and bright backgrounds. In a representative study, Wang \etal~\cite{wang2018traffic,wang2020traffic} applied HDR imaging to traffic light recognition for autonomous driving. In their approach, traffic light candidates were detected from the low-exposure frames and accurately classified using a DNN in the high-exposure frames. \cite{wang2020deep} presented a DNN method for detecting the source of HDR images from LDR images. 

Conversely, existing object detection networks trained
on general LDR images cannot detect relevant objects in adverse visual conditions.
\cite{mukherjee2020backward} proposed an approach to detect objects using HDR imaging under extreme lighting conditions based on existing CNN frameworks. The approach does not require HDR datasets and expensive retraining procedures because it uses HDR-to-LDR mapping techniques, which preserve the salient highlight, shadow, and color details. Existing DNN models use mapped LDR images to extract relevant features for object detection. \cite{wang2020multi} introduced a DNN-based framework for saliency detection of HDR content. The framework follows a similar idea as \cite{mukherjee2020backward}: first, HDR images are transformed to LDR images; next, neural networks detect the salient regions. Recently, \cite{onzon2021neural} proposed a joint learning approach in which exposure control is trained with object detection in an end-to-end manner.

\vspace{3pt}
\noindent \textbf{Semantic Segmentation.}
Semantic segmentation is an important per-pixel classification task. Recently, HDR imaging has been used to improve the performance of DNNs. \cite{weiher2019domain} proposed a preliminary method that leveraged HDR data to learn segmentation networks. It generates LDR images from HDR images using various one-mapping operators. The network can be trained using both LDR and HDR images. Domain adaptation is used for training synthetic and real-world datasets. \cite{satilmis2020per} presented a per-pixel classification method for identifying cloud types based on HDR sky images. Their method uses a patch-based feature extraction approach, which extracts the representative feature vectors used for labeling. 

\vspace{3pt}
\noindent \textbf{Stereo and Depth Estimation.}
As mentioned in Sec.~\ref{sec:stereo_hdr}, many attempts have been made to obtain HDR images from stereo LDR image pairs. This section analyzes some recent DL-based approaches that use HDR imaging for stereo matching and depth estimation.
The method in \cite{chen2020learning} is a representative study using multi-exposure LDR images for stereo matching, and the HDR fusion network is learned using the warped LDR images. \cite{im2018robust} introduced a depth estimation method from a short burst of multi-exposed LDR images. The estimated depth could align the LDR images, which are necessary for achieving HDR or image fusion.

\vspace{3pt}
\noindent \textbf{Illumination Estimation.}
Illumination is important for determining the appearance of a scene.  Correspondingly, research has focused on predicting HDR lighting conditions using LDR images. 
\cite{Wang_2019_CVPR,Zhang_2019_CVPR,Hold-Geoffroy_2017_CVPR,hold2019deep} are representative methods aiming to estimate the illumination in either indoor or outdoor environments using DNNs. The core pipeline of these methods is to train DNNs to achieve image-to-illumination mapping to model varying light conditions. However, learning illumination requires many high-quality datasets with the corresponding illumination conditions; thus, most methods explore indirect or multistep approaches to achieve this goal.  

\vspace{-5pt}
\section{Discussion and New Perspectives}
\label{sec:discussions}
\noindent \textbf{Pros and Cons of DL-based Methods vs. the Prior Arts.}
The common approach in the early methods \cite{sen2012robust,hu2013hdr,oh2014robust} are to generate HDR images by aligning multiple LDR images captured with different exposures via optimization. However, these methods have two distinct disadvantages. First, they failed to work on complex backgrounds and large motions. Second, optical flow-based alignment algorithms cannot generate new contents in the saturation and occlusion regions. By contrast, DL-based methods~\cite{wu2018deep,kalantari2017deep,yan2019attention} better address the artifacts induced in large foreground motions. Moreover, DNN-based optical flow algorithms can better align the LDR images and generate new content in the saturated and occlusion regions. Overall, the DL-based methods showed significant performance gains.

By contrast, traditional single-image HDR imaging methods estimate the density of light sources to expand the dynamic range \cite{akyuz2007hdr,huo2014physiological,kovaleski2014high,masia2017dynamic}. However, these methods cannot recover the brightness of the saturated regions as they do not utilize contexts in the images. Therefore, prior arts suffer from considerable artifacts and show less plausible performance, which can be verified in the experimental studies in the supplementary material. 
By contrast, DL-based methods \cite{eilertsen2017hdr,liu2020single, santos2020single} systematically utilize contextual information and produce better results in saturated regions. Although DL-based methods are promising, they are computationally more expensive, and it is necessary to create LDR scenes with diverse exposure information to reconstruct HDR images.  

\noindent \textbf{Exposure Bracket Selection for Deep HDR Imaging.}
Exposure bracket selection is an important factor in deep HDR imaging. This depends on several key factors. The first is the dynamic range of a scene. The exposure value (EV) determines exposure compensation, and is a term used for bracketing. Therefore, EV+0 is the exposure of the camera's default value. `EV-2' means that its exposure is `minus two shots'. EV+2 means that its exposure is `plus two shots'.
Three exposure shots at -2/0/+2 EV can usually handle most scenes, which is a common setting in representative works,~\eg, \cite{wu2018deep,yan2019attention,kalantari2017deep}. However, scenes with more over-exposed pixels, such as pixels with sunlight, may require more over-exposed LDR shots (\eg, 5 shots), and vice versa. Although more brackets contribute to better HDR imaging quality, the computation costs increase accordingly. Therefore, most deep HDR imaging methods consider three brackets, which are merged into a single HDR.

\noindent \textbf{Feature Representations in Network Design.}
Feature representation is an important factor in HDR imaging, especially in multi-exposure HDR image reconstruction. Learning features from LDR images using the attention mechanism are common; however, it requires more computations. A possible solution is to use more efficient yet effective attention modules, \eg,~\cite{chen2020dynamic}, to learn more correlated features. Another direction is to explore contrastive learning~\cite{chen2020simple} to learn better feature representations from LDR images. Moreover, the potential of GT HDR images has not yet been fully exploited. Existing methods only leverage HDR images for supervision; however, they can be used to learn privileged knowledge, as in \cite{lee2020learning}. Therefore, it is possible to add a removable network that encodes the feature representations of HDR images to guide the learning of LDR image features.

\vspace{3pt}
\noindent \textbf{Data-efficient Learning.}
A challenge in deep HDR imaging is the need for large-scale labeled datasets to train the DNN models. However, for multi-exposure HDR image reconstruction, building large-scale datasets is expensive and challenging. Therefore, it is necessary to explore more data-efficient deep HDR imaging methods.
One promising direction is to explore self-supervised or unsupervised learning to better explore visual information from unlabeled LDR images. In this situation, it is crucial to learn more reliable visual information based on data augmentation or adversarial examples \cite{yang2020adversarial}. Another possible solution is to explore data-free knowledge transfer \cite{chen2019data} to reduce the requirements for training data. Finally, the recent development of few-shot learning \cite{Sun_2019_CVPR} may be a way to learn robust HDR imaging networks, reducing the need for labeled data.

\vspace{3pt}
\noindent \textbf{Computationally Efficient Learning.}
Based on our review, the SoTA multi-exposure DHR image reconstruction methods include feature encoding, merging, and reconstruction. The encoding phase relies on learning to exclude misaligned features using attention or homography; the merging phase also needs to avoid harmful features. This inevitably leads to a large number of convolutional operations to ensure a robust performance. Consequently, DNNs are extremely deep and large. However, real-time methods are preferred in practice. Thus, there is an urgent need to develop lightweight HDR imaging models. One solution is to compress the learned models using network pruning \cite{liu2018rethinking} or knowledge distillation \cite{wang2021knowledge}. Another solution is to replace the standard convolution with depth-wise convolution or inverted residual structures \cite{sandler2018mobilenetv2}.

\vspace{3pt}
\noindent \textbf{Potential of Cross-Task Consistency.}
Research on HDR imaging with multitask learning has become more practical. To learn multiple tasks, existing methods mostly rely on sequential or joint learning. However, these methods incur considerable computational costs and require high-quality datasets. For instance, joint HDR and SR inevitably add the requirement for HR cameras to capture HR scenes. Recent progress in multitask learning~\cite{zamir2020robust} shows that the predictions of multiple tasks for the same image are not independent but are expected to be consistent. Therefore, future research may explore the intrinsic consistency between HDR imaging and other tasks. Importantly, new loss constraints are required to ensure cross-task consistency in HDR imaging.

\vspace{3pt}
\noindent \textbf{Potential of Novel Sensors.}
In Sec.~\ref{hdr_novelsensors}, we reviewed and analyzed deep HDR imaging techniques using recently developed sensors, such as event cameras and thermal sensors. These sensors usually capture scenes with a relatively high HDR. There have been some attempts to reconstruct HDR images using these sensors; however, existing methods should be further improved and studied. For instance, although the method in ~\cite{han2020neuromorphic} uses the reconstructed intensity maps to enhance the LDR-to-HDR image mapping, intensity generation is not differentiable; thus, the intensity maps cannot fully convey the HDR information of the event data. EvDistill~\cite{wang2021evdistill} is a method for joint learning of intensity reconstruction and end tasks in an end-to-end manner in LDR scenes. This might be a potential direction for event-guided HDR image reconstruction. Another solution might be to directly fuse the intermediate features of events and those of the LDR images using multimodality fusion techniques, such as \cite{joze2020mmtm}.

\vspace{3pt}
\noindent\textbf{Potential of GNNs for HDR Imaging.} Attention and non-local feature aggregation \cite{yan2020nonlocal,yan2019attention} have been shown to be effective for deep HDR imaging.  However, these methods exploit only the
similar patches of LDR images within the same scale. Although multiscale dense networks are important for learning the illuminance information in LDR images, they only aggregate the feature information in local regions. IGNN~\cite{zhou2020cross} adopts graph neural networks (GNNs), which aggregate similar cross-scale patches of the LR images in a graph,  to complement
external information learned from the training dataset. In HDR imaging, it has been shown that recurrence and correlation of local patches of LDR images are crucial for either exposure alignment or merging. Therefore, future research may explore the potential of GNNs to learn dense feature information from LDR images by exploring the cross-scale recurrence property in graphs.

\vspace{3pt}
\noindent \textbf{Toward Real-World HDR.}
Deep HDR imaging is a practical technique, but it is limited to real-world scenarios. For example, it suffers from unexpected saturated regions in the LDR images and limited exposure levels. Although many methods have achieved reasonable performance on simulated datasets, their performance dramatically drops on more complex real-world LDR scenes. Research should consider the lack of generalization of DNNs.
One solution is to apply unsupervised domain adaptation to adapt the simulated exposure information to DNN learning real-world LDR images. Another solution is to aggregate knowledge from multiple DNNs~\cite{wang2021knowledge} trained on various LDR contents and transfer the knowledge to the DNN learning real-world LDR images.

\vspace{-10pt}

\section{Conclusion}
We reviewed the major technical details and applications of deep HDR imaging. We formally defined the problem and introduced a hierarchical and structural taxonomy of deep HDR imaging methods. For each category in the taxonomy, we reviewed the current technical status, potential, and challenges. Drawing connections among these approaches highlights new active research fields and is likely to identify new research problems and methods by exploring the potential of each paradigm. Based on our review and analysis, we discussed the ways in which the existing challenges can be overcome and the research gaps can be addressed.

\ifCLASSOPTIONcompsoc
  \section*{Acknowledgments}
\else
  \section*{Acknowledgment}
\fi

This work was supported by the National Research Foundation of Korea (NRF) grant funded by the Korea government (MSIT) (NRF-2018R1A2B3008640).

\bibliographystyle{IEEEtran}
\bibliography{reference}

    
    

\ifarXiv
    

\section*{Supplementary material}

# Deep Learning for HDR Imaging:
# State-of-the-Art and Future Trends
# –Appendix–

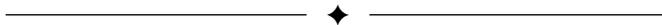

## 1 DETAILS OF DATASET FOR DEEP HDR IMAGING

The dataset is important for the success of deep HDR imaging methods. Various benchmark datasets for deep HDR imaging differ in exposure levels, dataset sizes, image resolution, quality, scene diversity, etc. Some datasets obtain LDR images by simulating the CRF or using a CRF database, while others contain real-world LDR images with their ground-truth (GT) HDR images. Table 1 of the main paper summarizes commonly used benchmark datasets for deep HDR imaging, including the data size, data type, availability of the GT, spatial resolution, and scene details. In addition, some datasets have been developed for multiple-exposure fusion (MEF) [1], [2], video HDR [3], [4], and HDR imaging with novel sensors [5], [6].

Regarding the synthetic HDR datasets, the HDR GT images are collected from diverse real-world scenes. For single HDR reconstruction, the most common approach is to imitate the LDR formation pipeline using different CRF functions from a set of CRF databases [7], [8], or a virtual camera to capture a number of random regions of the scene based on randomly selected camera curves [9], [10], [11]. For the real-world singe image HDR dataset in, *e.g.*, [7], 600 amateurs are instructed to capture scenes with multiple exposures from a steady tripod. The LDR exposure stacks are fused to obtain HDR images using Photomatix [1].

Regarding the multi-exposure HDR imaging, the dataset collection process from [12] is the representative. This dataset aims to obtain a set of LDR images of dynamic scenes and their corresponding ground truth HDR image. In the data generation process of the dynamic scene, it was assumed that the scene was static, and the human subject was moving to simulate motion between the LDR images. Note that the HDR image was aligned with the reference image (middle exposure). With the reference LDR images, the up-and down-exposed LDR images are captured by asking the subject to move by simulating camera motion or no motion. The HDR ground truth images were generated by capturing a set of static LDR images (the human subject kept still) at different exposure values on a tripod. In such a way, the dynamic set consists of the down-exposed and up-exposed images and the inference image. For the static scene, the HDR image generation is similar to the one in [7].

1. https://www.hdrsoft.com/

## 2 DETAILS OF OPTIMIZATION AND LOSS FUNCTIONS

### 2.1 Loss functions

**Pixelwise Loss.** Pixelwise loss is a simple yet common loss used in deep HDR reconstruction, which is usually based on L1 or L2 norm. This loss usually measures the pixel-wise difference between the reconstructed tone-mapped HDR image $\hat{H}$ and the GT HDR image $H$, which can be formulated as follows:

$$\mathcal{L}_{pw} = ||T(H) - T(\hat{H})||_p, \qquad (1)$$

where $p$ is the norm (1 or 2), and $T$ represents the tone-mapping operator based on the $\mu$-law defined in Eq. 2:

$$T(H) = \frac{\log(1 + \mu H)}{\log(1 + \mu)}, \qquad (2)$$

where $\mu$ denotes the amount of compression. Note that a more naive approach is to directly calculate the error between the GT $H$ and $\hat{H}$, which leads to a smaller error estimation in the linear HDR domain. However, the estimated HDR images after tone mapping are often degraded by noise and artifacts.

**SSIM Loss.** As mentioned in Sec.2 of the main paper, SSIM measures the structural difference between the reconstructed HDR and the GT images. SSIM served as a common objective function in deep HDR imaging, such as [13], [14]. Other studies used the multiscale SSIM (MS-SSIM) [15] and MEF-SSIM [16] as the loss functions for multi-exposure HDR image reconstruction.

**Perceptual Loss.** Pixelwise loss usually leads to blurring of reconstructed HDR images. To improve perceptual quality, the perceptual loss has been used in many studies, *e.g.*, [1], [10], [17], [18]. The perceptual loss essentially transfers the extracted knowledge from the hierarchical image features from the classification network, *e.g.*, VGG [19], to HDR imaging networks. Perceptual loss encourages the features of the reconstructed HDR images to be perceptually similar to those of the GT HDR images. Mathematically, this loss can be formulated as follows:

$$\mathcal{L}_{Perceptual} = \sum_l ||\phi_l(T(H) - \phi_l(T(\hat{H}))||_1, \qquad (3)$$

where $\phi_l$ is the feature map extracted from the $l$th layer of the classification network.

**GAN Loss.** GAN [20], [21] has been broadly used in deep HDR imaging to improve the learning quality or reduce the need for labeled data. The GAN consists of a generator and discriminator, aiming to distinguish the generated HDR image from the real one.

TABLE 1: The effectiveness of HDR image reconstruction when using some representative loss functions.

| Loss | PSNR | SSIM |
|------|------|------|
| $L_1$ | 13.54 | 0.7053 |
| $L_2$ | 13.68 | 0.7059 |
| $L_1$ + Perceptual loss | 14.71 | 0.7208 |
| $L_1$ + Cosine loss | 14.53 | 0.7196 |
| $L_1$ + SSIM loss | 14.51 | 0.7194 |
| $L_1$ + Style loss | 14.60 | 0.7197 |
| $L_1$ + Gradient loss | 14.37 | 0.7097 |

The HDR network, regarded as the generator, is trained alternately with the discriminator during the training. Many studies on HDR imaging have proposed the use of adversarial loss to facilitate learning. For instance, Niu *et al.* [22] proposed HDR-GAN to learn ghost-free HDR imaging under large motions. The GAN loss is based on SphereGAN [23]. Other studies, *e.g.*, [2], [11], used LSGAN [24] and conventional GAN [20] loss, respectively. We discuss GAN-based HDR imaging in Sec.6.2 of the main paper.

**Gradient Loss.** The gradient loss computes the gradient magnitude of the tone-mapped HDR image to retain the high-frequency content. Therefore, the addition of this loss constrains the gradient difference, which slightly improves the performance and makes the fused image exhibit a sharper appearance [2], [25], [26].

**Cycle Loss.** Cycle consistency is broadly used in image translation frameworks [27]. In [28], an LDR sub-network was proposed to recover three static LDR images corresponding to the reconstructed HDR image. This adds constraint for LDR-to-HDR mapping, reducing the ghosting effects of multi-exposure HDR image reconstruction.

**Other Losses.** In [10], style loss captured style and texture by comparing digital statistics with the Gram matrix collected over the entire HDR image. Moreover, the cosine similarity loss [29] ensured that each pixel pointed in the same direction of the three-dimensional RGB space. It provided improved color stability, especially for low luminance values in HDR images. In addition, some studies [30], [31] proposed adding regularization to pixel-wise loss to ensure the stability of gradient descent.

## 2.2 Experimental validations

In this section, we conduct experiments to validate the effectiveness of the commonly used loss functions. We take FHDR [17] as the baseline network model for the experiments. We use the 'HDR-Real' dataset [7] for training and the test set of [17] for evaluation in all experiments. For a fair comparison of loss functions, we train FHDR for 20 epochs, which we find is enough for convergence. The $l_1$ loss is the most commonly used loss function, and we use it as the baseline loss. We then change to $l_2$ loss to see how the quality of HDR image reconstruction changes. In addition, we add additional loss terms on the $l_1$ loss to study how the HDR image reconstruction quality varies. The objective function can be formulated as:

$$L = L_1 + \lambda L_{other} \quad (4)$$

Regarding the perceptual loss, we set the parameter $\lambda$ as 10. For the cosine similarity loss, gradient loss, style loss, and SSIM loss, we set the parameter $\lambda$ as 0.1 for convenience. When it comes to GAN loss, we set $\lambda$ 0.01. We use the PSNR and SSIM to measure the HDR image reconstruction quality. The numerical results of HDR image reconstruction using different loss functions are shown in Table. 1.

As shown from Table. 1, $L_1$ and $L_2$ are the most commonly used pixel-wise matching loss for deep HDR imaging. Both loss terms are effective although $L_2$ shows a litter higher PSNR scores than that of $L_1$. Please be noted that there is no absolute standard of choosing $L_1$ or $L_2$ loss. For instance, in some representative methods, *e.g.*, [8], [10], [11], [26], [32], [33], [34], [35], $L_1$ loss is mainly used. In contrast, other, *e.g.*, [9], [15], [31], [36], [37], [38], [39], [40] use $L_2$ loss.

On the other hand, when additional loss terms are added based on the pixel-wise loss, according to Eq. 4, the HDR image reconstruction performance are enhanced accordingly. In particular, perceptual loss and style loss are targeted to improve the perceptual quality of HDR image reconstruction. Adding perceptual loss improves PSNR scores by around $1.2dB$. Meanwhile, adding cosine similarity loss improves HDR image reconstruction quality, with around $0.9dB$ increment of PSNR scores. Moreover, adding gradient loss is slightly less effective than adding the perceptual loss, with the PSNR gap of $0.3dB$.

**Discussion.** We have compared the effectiveness of adding different loss terms for deep HDR imaging based on the experiments. The pixel-wise loss is the fundamental yet effective loss for deep HDR imaging. Adding other loss terms, *e.g.*, the perceptual loss, also improves the HDR image reconstruction quality. However, the capacity of these loss functions is still limited. Moreover, they are also commonly used in other low-level vision problems, such as image super-resolution. In other words, they are less tailored for HDR imaging problems. Therefore, future research could consider the specialty of HDR imaging and propose more effective loss functions.

## 3 DISCUSSION AND CHALLENGES FOR HDR IMAGING WITH SR

We have analyzed the representative deep HDR imaging with SR in the Sec.4.1 of the main paper. We now discuss the issues and challenges of the representative methods. As summarized in Table 2, joint learning is a more efficient method for HDR imaging with SR. However, most methods focus on single-exposure HDR imaging, following the SISR paradigm. Although [41] proposed a GAN-based architecture to apply joint HDR-SR learning, it remains challenging to capture high-frequency textures while enhancing the local contrast. Furthermore, all methods adopt the "ideal" degradation kernel, namely bicubic sampling, to obtain LR LDR images. However, as studied in SISR, real-world LR images are obtained with unknown degradations [42], [43]. Therefore, future research should estimate degradation while enhancing the local contrast. One possible way might be combining the bind SR methods, *e.g.*, [44], [45], [46], [47], with reference-based HDR imaging methods. Additionally, all SoTA methods are supervised and require a large amount of labeled data; however, high-quality datasets are limited. Therefore, future directions might explore, *e.g.*, few-shot learning [48] or knowledge transfer [49], to reduce the need for labeled data.

## 4 COMPARISON OF THE REPRESENTATIVE DEEP HDR IMAGING METHODS

### 4.1 Deep Multi-exposure HDR Methods

#### 4.1.1 Experiment Settings and Datasets

We have tried to compare the representative methods listed in Table 2 of the main paper for multi-exposure methods. However,

TABLE 2: Deep HDR imaging and super-resolution employed by some representative methods.

| Method | Publication | No. of Input | Learning strategy | Alignment | Fusion | Upsample method | Image decomposition | Loss |
|--------|-------------|--------------|-------------------|-----------|--------|-----------------|---------------------|------|
| Kim [36] | ACCV'18 | 1 | Joint | ✗ | Feature | Pixel Shuffle | ✗ | Joint + SR + HDR |
| Kim [50] | ICCV'19 | 1 | Joint | ✗ | Feature | Pixel Shuffle | Luminance + Reflection | L2 |
| Kim [51] | AAAI'20 | 1 | Joint | ✗ | Feature + Image | Pixel Shuffle | Luminance + Reflection | L2 + GAN |
| Ashwath [52] | ECCV'20 | 3 | Sequential | Optical flow | Image | Guided Upsampling | ✗ | BCE + L2 + L2 |
| Deng [53] | IEEE TIP'21 | 2 | Joint | ✗ | Feature + Image | Deconvolution | ✗ | MS-SSIM |

TABLE 3: Comparison of multi-exposure HDR imaging performance, model parameters, and computation costs based on some representative methods. '−' indicates 'not available' in some methods.

| Algorithm type | Method | PSNR-$\mu$ | PSNR-L | HDR-VDP-2 | SSIM-$\mu$ | SSIM-L | Run time (s) | Parameters (M) |
|----------------|--------|-----------|--------|-----------|-----------|--------|--------------|----------------|
| Patch-based | Sen [54] | 40.9453 | 38.3147 | 55.7240 | 0.9805 | 0.9726 | 61.81 (CPU) | − |
| | Hu [55] | 32.1872 | 30.8395 | 55.2496 | 0.9716 | 0.9506 | 79.77 (CPU) | − |
| | Oh [56] | 27.3510 | 27.1119 | 46.8259 | 0.9042 | 0.8857 | 65.15 (CPU) | − |
| DL-based | Endo [8] | 14.0925 | 13.1116 | 47.7399 | 0.7224 | 0.6524 | 1.52 | 48.11 |
| | Eilertsen [9] | 8.2123 | 8.6846 | 44.3944 | 0.7409 | 0.6658 | 0.25 | 29.44 |
| | Kalantari [12] | 42.7423 | 41.2158 | 60.5088 | 0.9877 | 0.9848 | 0.34 | 0.3 |
| | Yan [57] | 42.2263 | 41.0170 | 60.2991 | − | − | − | − |
| | Wu [37] | 41.6377 | 40.9082 | 60.4955 | 0.9869 | 0.9858 | 0.22 | 20.4 |
| | Yan [34] | 43.6310 | 41.1430 | 61.1040 | 0.9900 | 0.9702 | 0.30 | 1.5 |
| | Yan [38] | 41.4143 | 41.4255 | 61.2107 | 0.9887 | 0.9857 | 0.31 | 38.1 |
| | Choi [35] | 43.2120 | 41.6970 | 62.4810 | − | − | 0.59 | − |
| | Pu [26] | 43.8487 | 41.6452 | 62.5495 | 0.9906 | 0.9870 | − | 5.01 |
| | Yan [28] | 42.8768 | 41.3013 | 60.6421 | 0.9882 | 0.9852 | − | 4.60 |
| | Niu [22] | 43.9220 | 41.5720 | 65.4500 | 0.9905 | 0.9865 | 0.17 | 2.56 |

as the source codes of some methods are not available, we emailed the authors to inquire about some visual results and the model parameters and run-time. For some methods, we directly utilize the officially reported results. We mainly used the benchmark dataset built by Kalantari *et al.* [12] for comparison and validation. It is also possible to use other HDR datasets; however, most do not provide ground truth HDR images or have limited scenes. The dataset of [12] contains 89 scenes with the GT HDR images. In this dataset, each scene was taken with three different exposure shots when the object moved. Meanwhile, three other shots were taken with no object movement. The aim is to generate a GT HDR image based on the reference LDR image, namely, the medium exposed LDR image. All the images in this dataset were resized to 1000x1500. We use three evaluation metrics: PSNR, SSIM, and HDR-VDP-2. The PSNR and SSIM values are calculated between the generated HDR and the GT HDR image after tone-mapping using $\mu$-law (PSNR-$\mu$, SSIM-$\mu$) and linear domains (PSNR-L, SSIM-L). The HDR-VDP2 scores are directly obtained from the generated linear HDR image and the GT HDR image.

#### 4.1.2 Experimental Analysis

The quantitative results, accompanied by the qualitative results of the representative methods, are provided in Table 3 and Figure 1 due to the limitation of space in the main paper. To better validate the effectiveness of the deep learning-based HDR imaging methods, we also selected several non-deep learning-based (patch-based) methods, such as Sen *et al.*. [54], Hu *et al.* [55], and Oh *et al.* [56] for comparison.

First of all, compared with the patch-based methods, the deep learning-based methods show significant performance gains. For instance, compared with the SoTA patch-based method, Sen *et al.*, the HDR-VDP-2 score difference with the SoTA deep learning-based method, Niu *et al.* [22], is around 10, which is indeed a large difference. Qualitatively, as shown in Fig.1, the patch-based methods can not recover the details in the saturated regions. Moreover, we can see that the patch-based methods suffer from noticeable artifacts due to the misalignment of multiple exposure images.

Furthermore, the deep single HDR imaging method, *e.g.*, Endo *et al.* [8] and Eilertsen *et al.* [9], only use a single LDR image. Although they can reduce the ghosting artifacts, these methods can not produce sharp results and suffer from color distortions due to the highly ill-posed mapping from LDR-to-HDR. As a consequence, the numerical scores, such as HDR-VDP-2 scores, are significantly lower than the deep multi-exposure HDR imaging methods, such as Niu *et al.* [22].

Last but not least, the multi-exposure HDR imaging method, Kalantari *et al.* [12], produces visible artifacts, as shown in the second column of Fig.1. This is mainly caused by two possible reasons. The first reason is the misalignment of optical flow in large foreground motions. The second one is the less accurate merging method. The method [37] generates better HDR images than [12]; however, it smooths out the texture details and can not fully remove the ghosting effects. In contrast, Yan *et al.* [38], [52] suppresses the ghosting artifacts and recovers the saturated regions. Niu *et al.* [22] outperforms all the other methods in the metric of HDR-VDP-2, which can also be visually verified in the qualitative results in Fig.1. Overall, the method proposed in Niu *et al.*produces better HDR images and demonstrates a noticeable margin than the other methods.

## 4.2 Deep single image HDR methods

### 4.2.1 Experiment Settings and Datasets

We have also tried to compare with the representative deep single image HDR methods, as originally summarized and analyzed in Table.4 of the main paper. As the source codes of some methods are not available, we again emailed the authors to inquire about some visual results and the model parameters and run-time. For some methods, we directly take the results reported in their papers.

We conducted experiments on the two commonly used benchmark datasets, the RAISE, and HDR-EYE datasets, for evaluation and comparison. We provide both qualitative and quantitative comparisons for the representative methods on these datasets in this revision. To better validate the effectiveness of the deep learning-based HDR imaging methods, we again selected several non-deep learning-based methods, such as AEO [58], HPEO [59], KOEO [60] and MEO [61] for comparison. We mainly used the mean and standard deviation of HDR-VDP-2 as the evaluation metric used in the representative methods. Please note that we compared the linearly mapped HDR images with the ground truth images using HDR-VDP-2.

### 4.2.2 Experimental results

The experimental studies of the representative methods for deep single image HDR imaging are provided in Table.5 and Figure 2 due to the limitation of space in the main paper.

The quantitative results are shown in Table 5. Compared with the SoTA non-deep learning-based method, such as AEO, the SoTA deep learning methods, such as Kim *et al*. [62] does show significant performance gains, with an increase of HDR-VDP-2 score of 4. Please be noted that this is a significant enhancement of HDR quality for single HDR imaging. This is because single image HDR imaging is more challenging as it is an ill-posed problem and has limited exposure information in a single image. On the other hand, when comparing the deep single image HDR imaging methods, we find that learning the LDR-to-HDR mapping is important. For instance, the methods [8], [11], [63] are focused on generating the synthetic LDR brackets from a single HDR image. The generated LDR image brackets are merged into an HDR image using the tone-mapping database or functions. However, the performance gains are still limited. This is because the mapping of LDR-to-HDR is not end-to-end mapping, and the training to generate virtual LDR brackets from the training data does not generalize well to the test LDR images. This important factor was discussed and validated in [62], where the authors proposed an end-to-end approach to recursively learn to generate LDR brackets and simultaneously synthesize the HDR images with the correctly estimated camera response function (CRF).

Although Liu *et al*.. [7] did not focus on generating LDR brackets, it achieved comparable performance with [62], as shown in Table 4. The reason is that it better modeled the LDR image formation pipeline and accordingly proposed a deep learning framework to learn the HDR-to-LDR image formation process. Therefore, it performed favorably against the SoTA single image HDR imaging methods. The qualitative results in Fig.2 both show the visual results, consistent with Table.5. For instance, in Fig.2, the trees in the arch area reconstructed by the non-deep learning methods suffer from noticeable artifacts and wrongly recovered exposure information. In comparison, the SoTA deep learning methods such as [7], [62] reconstruct better HDR images, which can be validated from the over-exposed regions in the LDR images.

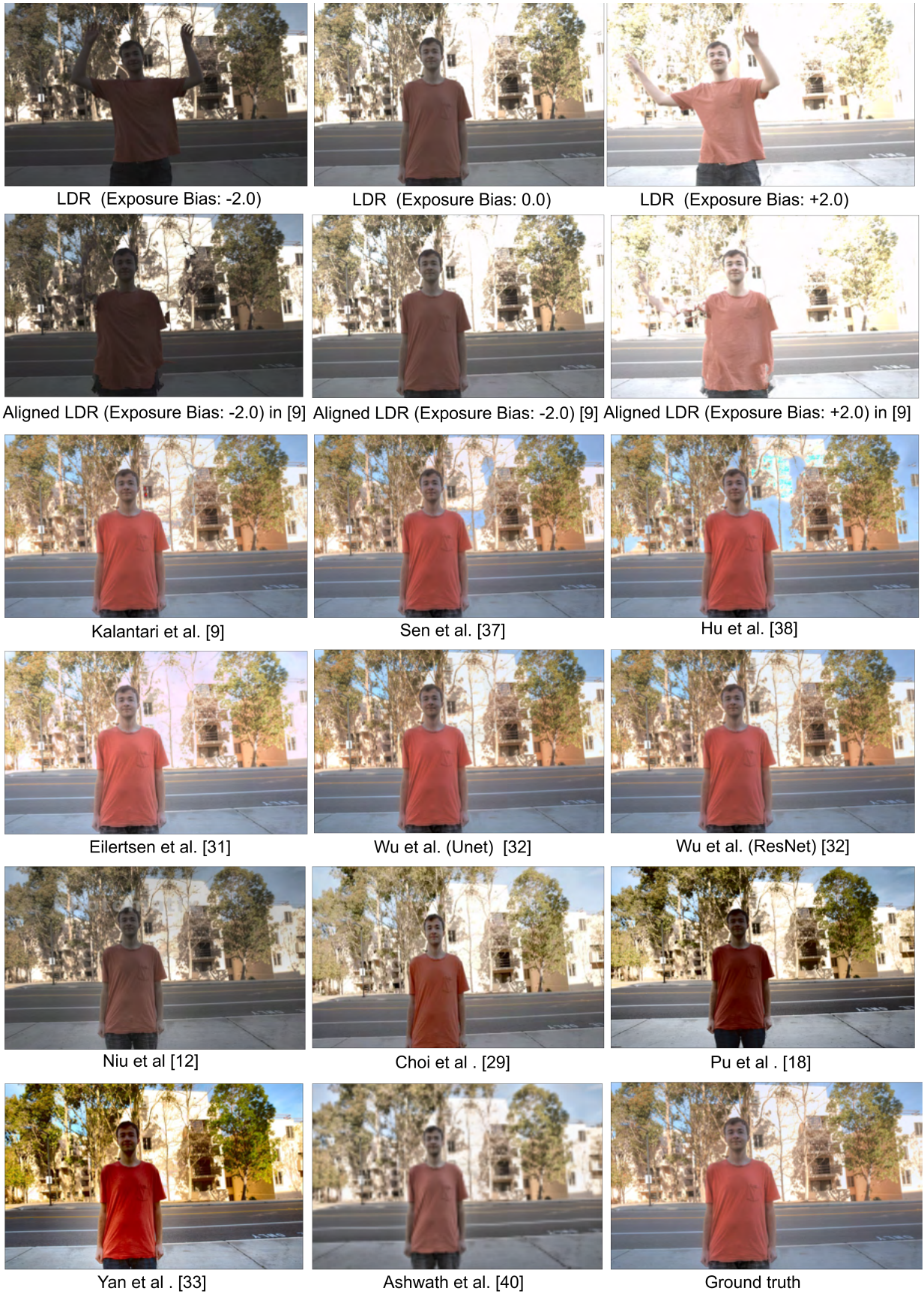

Fig. 1: Visual results on the representative deep multi-exposure HDR imaging methods.

TABLE 4: Comparison of single-exposure HDR imaging performance, model parameters, and computation costs based on some representative methods. '−' indicates 'not available' in some methods.

| Algorithm type | Method | RAISE dataset | | HDR-EYE dataset | | Run time | Parameters |
|---|---|---|---|---|---|---|---|
| – | | Mean | STD | Mean | STD | Seconds (s) | Million (M) |
| Non-DL | AEO [58] | 56.54 | 4.00 | 49.20 | 5.81 | – | – |
| | HPEO [59] | 55.16 | 5.40 | 44.12 | 4.75 | – | – |
| | KOEO [60] | 54.33 | 4.72 | 49.62 | 6.06 | – | – |
| | MEO [61] | 56.30 | 3.72 | 49.28 | 6.09 | – | – |
| DL | Eilertsen et al. [9] | 57.08 | 3.77 | 50.80 | 56.79 | 0.334 | 29.44 |
| | Endo et al. [8] | 57.69 | 4.08 | 51.80 | 5.93 | 1.489 | 48.09 |
| | ExpandNet [29] | 54.62 | 1.99 | 50.43 | 5.49 | 0.063 | 0.45 |
| | Deep chain HDRI [11] | – | – | 49.80 | 5.97 | – | – |
| | Deep recursive HDRI [63] | 57.57 | 3.70 | 48.85 | 4.91 | 0.61 | 6.97 |
| | FHDR [17] | 59.14 | 2.76 | 53.81 | 3.6 | 0.22 | 0.57 |
| | Kim et al. [62] | 59.49 | 3.42 | 55.91 | 1.92 | 0.96 | 232 |
| | SingleHDR [7] | 59.30 | 3.54 | 54.51 | 3.71 | 0.52 | 29.01 |

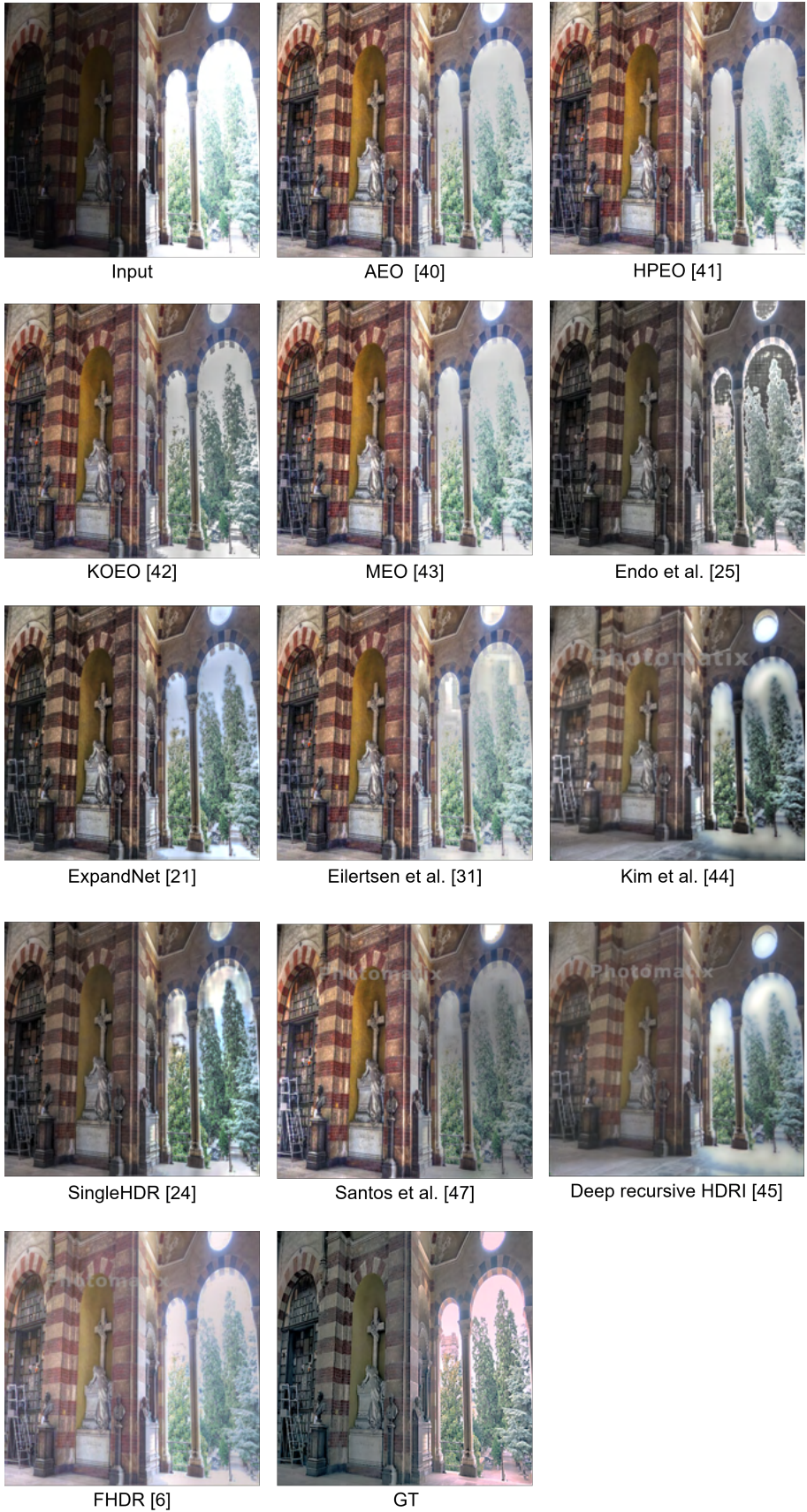

Fig. 2: Visual results on the representative deep single HDR imaging methods.


\fi

\end{document}